\let\includefigures=\iftrue
%
%
\let\useblackboard=\iftrue
%
%
\newfam\black
\input harvmac.tex
\includefigures
\message{If you do not have epsf.tex (to include figures),}
\message{change the option at the top of the tex file.}
\input epsf
\def\figin{\epsfcheck\figin}\def\figins{\epsfcheck\figins}
\def\epsfcheck{\ifx\epsfbox\UnDeFiNeD
\message{(NO epsf.tex, FIGURES WILL BE IGNORED)}
\gdef\figin##1{\vskip2in}\gdef\figins##1{\hskip.5in}
\else\message{(FIGURES WILL BE INCLUDED)}%
\gdef\figin##1{##1}\gdef\figins##1{##1}\fi}
\def\DefWarn#1{}
\def\figinsert{\goodbreak\midinsert}
\def\ifig#1#2#3{\DefWarn#1\xdef#1{fig.~\the\figno}
\writedef{#1\leftbracket fig.\noexpand~\the\figno}%
\figinsert\figin{\centerline{#3}}\medskip\centerline{\vbox{\baselineskip12pt
\advance\hsize by -1truein\noindent\footnotefont{\bf Fig.~\the\figno:} #2}}
\bigskip\endinsert\global\advance\figno by1}
\else
\def\ifig#1#2#3{\xdef#1{fig.~\the\figno}
\writedef{#1\leftbracket fig.\noexpand~\the\figno}%
\global\advance\figno by1}
\fi
\def\Title#1#2{\rightline{#1}
\ifx\answ\bigans\nopagenumbers\pageno0\vskip1in%
\baselineskip 15pt plus 1pt minus 1pt
\else
\def\listrefs{\footatend\vskip 1in\immediate\closeout\rfile\writestoppt
\baselineskip=14pt\centerline{{\bf References}}\bigskip{\frenchspacing%
\parindent=20pt\escapechar=` \input
refs.tmp\vfill\eject}\nonfrenchspacing}
\pageno1\vskip.8in\fi \centerline{\titlefont #2}\vskip .5in}

\ifx\answ\bigans\def\tcbreak#1{}\else\def\tcbreak#1{\cr&{#1}}\fi
\useblackboard
\message{If you do not have msbm (blackboard bold) fonts,}
\message{change the option at the top of the tex file.}
\font\blackboard=msbm10 scaled \magstep1
\font\blackboards=msbm7
\font\blackboardss=msbm5
\textfont\black=\blackboard
\scriptfont\black=\blackboards
\scriptscriptfont\black=\blackboardss
\def\Bbb#1{{\fam\black\relax#1}}
\else
\def\Bbb{\bf}
\fi
%
\def\yboxit#1#2{\vbox{\hrule height #1 \hbox{\vrule width #1
\vbox{#2}\vrule width #1 }\hrule height #1 }}
\def\fillbox#1{\hbox to #1{\vbox to #1{\vfil}\hfil}}
\def\ybox{{\lower 1.3pt \yboxit{0.4pt}{\fillbox{8pt}}\hskip-0.2pt}}

\def\comments#1{}
\def\cc{{\rm c.c.}}

\def\QM{\Bbb{M}}

\def\QX{\Bbb{X}}
\def\QZ{\Bbb{Z}}
\def\p{\partial}

\def\eps{\epsilon}
\def\half{{1\over 2}}
\def\Tr{{{\rm Tr\  }}}
\def\tr{{\rm tr\ }}
\def\Re{{\rm Re\hskip0.1em}}
\def\Im{{\rm Im\hskip0.1em}}

\def\ket#1{|#1\rangle}
\def\vev#1{\langle{#1}\rangle}

\def\CA{{\cal A}}

\def\CF{{\cal F}}

\def\CM{{\cal M}}
\def\CN{{\cal N}}

\def\CS{{\cal S}}

\def\nl{\hfill\break}

\def\ap{\alpha'}

\def\I{I}

\def\II{\relax{I\kern-.10em I}}
\def\IIa{{\II}a}
\def\IIb{{\II}b}
\def\zo{z^1}
\def\zt{z^2}
\def\zbo{\bar z^{\bar 1}}
\def\zbt{\bar z^{\bar 2}}
\def\po{\psi^1}
\def\pt{\psi^2}
\def\pbo{\bar \psi^{\bar 1}}
\def\pbt{\bar \psi^{\bar 2}}

\def\tpo{\tilde\psi^1}
\def\tpt{\tilde\psi^2}

\def\cone{{\bf 1}}
\def\cI{{\bf I}}
\def\cJ{{\bf J}}
\def\cK{{\bf K}}

\def\Hom{{\rm Hom}}
\def\hk{hyperk\"ahler\  }
\def\Hk{Hyperk\"ahler\  }

\def\IZ{\relax\ifmmode\mathchoice
{\hbox{\cmss Z\kern-.4em Z}}{\hbox{\cmss Z\kern-.4em Z}}
{\lower.9pt\hbox{\cmsss Z\kern-.4em Z}}
{\lower1.2pt\hbox{\cmsss Z\kern-.4em Z}}\else{\cmss Z\kern-.4em
Z}\fi}
\def\IB{\relax{\rm I\kern-.18em B}}
\def\IC{{\relax\hbox{$\inbar\kern-.3em{\rm C}$}}}
\def\ID{\relax{\rm I\kern-.18em D}}
\def\IE{\relax{\rm I\kern-.18em E}}
\def\IF{\relax{\rm I\kern-.18em F}}
\def\IG{\relax\hbox{$\inbar\kern-.3em{\rm G}$}}
\def\IGa{\relax\hbox{${\rm I}\kern-.18em\Gamma$}}
\def\IH{\relax{\rm I\kern-.18em H}}
\def\II{\relax{\rm I\kern-.18em I}}
\def\IK{\relax{\rm I\kern-.18em K}}
\def\IP{\relax{\rm I\kern-.18em P}}

\def\lieg{{\underline{\bf g}}}

\def\inbar{\,\vrule height1.5ex width.4pt depth0pt}
\def\mod{{\rm mod}}

\def\p{\partial}

\font\cmss=cmss10 \font\cmsss=cmss10 at 7pt
\def\IR{\relax{\rm I\kern-.18em R}}

\def\qmvw{\CM_{\vec \zeta}(\vec v, \vec w) }

\def\Tr{\rm Tr}

\def\wb{{\bar{w}}}

\def\zb {{\bar{z}}}
\def\BR{\IR}
\def\BZ{\IZ}
\def\BR{\IR}
\def\BC{\IC}
\def\BM{\QM}

\def\BX{\QX}
\Title{ \vbox{\baselineskip12pt\hbox{hep-th/9603167}
\hbox{RU-96-15}
\hbox{YCTP-P5 -96 }}}
{\vbox{
\centerline{D-branes, Quivers, and ALE Instantons}}}
\centerline{Michael R. Douglas}
\smallskip
\smallskip
\centerline{Department of Physics and Astronomy}
\centerline{Rutgers University }
\centerline{Piscataway, NJ 08855-0849}
\centerline{\tt mrd@physics.rutgers.edu}
\smallskip
\centerline{\it and}
\smallskip
\centerline{Gregory Moore}
\smallskip
\centerline{Department of Physics}
\centerline{Yale University }
\centerline{New Haven, CT }
\centerline{\tt moore@castalia.physics.yale.edu}
\bigskip
\bigskip
\noindent
Effective field theories in type \I\ and \II\
superstring theories
for  D-branes located at points in the orbifold $\BC^2/\BZ_n$
  are supersymmetric gauge theories whose
field content is conveniently summarized by a `quiver diagram,'
and whose Lagrangian includes non-metric couplings to the orbifold moduli:
in particular, twisted sector moduli couple
as Fayet-Iliopoulos terms in the gauge theory.

\noindent
These theories describe D-branes on {\it resolved}
ALE spaces.  Their spaces of vacua are
moduli spaces of smooth ALE metrics and Yang-Mills instantons,
whose metrics are explicitly computable.
For $U(N)$ instantons, the construction exactly reproduces results of
Kronheimer and Nakajima.

\Date{March 19, 1996}
%
\lref\dlp{J.~Dai, R.~G.~Leigh and J.~Polchinski, Mod. Phys. Lett. {\bf A4}
(1989) 2073.}
\lref\pol{J.~Polchinski, Phys.~Rev.~Lett.~75 (1995) 4724-4727;
hep-th/9510017.}
\lref\chs{C. Callan, J. Harvey and A. Strominger, Nucl. Phys.
{\bf B367} (1991) 60.}
\lref\witten{E. Witten, ``Small Instantons in String Theory,'' hep-th/9511030.}
\lref\kn{P.B. Kronheimer and H. Nakajima, ``Yang-Mills instantons
on ALE gravitational instantons,'' Math. Ann. {\bf 288} (1990) 263.}
\lref\gimon{E. G. Gimon and J. Polchinski,
``Consistency Conditions for Orientifolds and D-manifolds,'' hep-th/9601038.}
\lref\hst{P. Howe, G.Sierra and P. Townsend, Nucl. Phys. {\bf B221} (1983)
331;\nl
G. Sierra and P. Townsend,
``The gauge invariant N=2 supersymmetric
sigma model with general scalar potential,''  Nucl. Phys. B233 (1984) 289.}
\lref\douglas{M.~R.~Douglas, ``Branes within Branes,'' hep-th/9512077.}
\lref\li{M.~Li, ``Boundary States of D-Branes and Dy-Strings,''
hep-th/9510161.}
\lref\call{C. G. Callan, C. Lovelace, C. R. Nappi and S. A. Yost,
	Nucl. Phys. {\bf B308} (1988) 221.}
\lref\polcai{J.~Polchinski and Y.~Cai, Nucl.~Phys. {\bf B296} (1988) 91.}
\lref\dsw{M. Dine, N. Seiberg and E. Witten,
	Nucl.~Phys. {\bf B289} (1987) 589.}
\lref\dis{M. Dine, I. Ichinose and N. Seiberg,
	Nucl.~Phys. {\bf B293} (1987) 253.}
\lref\superspace{{\it Superspace}, S. J. Gates, M. T. Grisaru,
M. Rocek and W. Siegel, Benjamin-Cummings 1983,
pp. 186-193 and 223-227.}
\lref\grojn{I. Grojnowski,
`` Instantons and affine algebras I: The Hilbert schemeand vertex operators,''
alg-geom/9506020}
\lref\hitchin{N.~Hitchin, A.~Karlhede, U.~Lindstr\"om, and M.~Ro{\v c}ek,
	Comm.~Math.~Phys. 108 (1987) 535.}
\lref\strom{A. Strominger, Nucl. Phys. B451 (1995) 96-108; hep-th/9504090.}
\lref\joerev{S. Chaudhuri, C. Johnson, and J. Polchinski,
``Notes on D-Branes,'' hep-th/9602052.}
\lref\kronheimer{P. Kronheimer, ``The construction of ALE spaces as
hyper-kahler quotients,'' J. Diff. Geom. {\bf 28}1989)665;
``A Torelli-Type Theorem for Gravitational Instantons,''
J. Diff. Geom. {\bf 29} (1989) 685}
\lref\kricm{P. Kronheimer, contribution to the 1990
ICM proceedings.}
\lref\vw{C. Vafa and E. Witten,
``A Strong Coupling Test of S-Duality,''
hep-th/9408074}
\lref\lmns{A. Losev, G. Moore, N. Nekrasov,
and S. Shatashvili, ``Four-dimensional Avatars of 2D CFT,''
hep-th/9509151}.
\lref\dmns{M. Douglas, G. Moore, N. Nekrasov, and
S. Shatashvili, ``D-branes and the construction of
$SO(w)$ instantons on ALE spaces,''  Yale preprint YCTP-P6-96,
 to appear}
\lref\hitchiniii{
N. Hitchin, ``Polygons and gravitons,''
Math. Proc. Camb. Phil. Soc, (1979){\bf 85} 465}
\lref\hitchini{N. Hitchin, ``Hyperk\"ahler Manifolds,''
Sem. Bourbaki, Asterisque 206 (1992)137}
\lref\nakalg{H. Nakajima, ``Instantons on ALE spaces,
quiver varieties, and Kac-Moody algebras,'' Duke Math. {\bf 76}(1994)365-416;
``Instantons and affine Lie algebras,''  alg-geom/9510003}
\lref\nakresol{H. Nakajima, ``Resolutions of
moduli spaces of ideal instantons on $\IR^4$,''
in {\it Topology, Geometry, and Field Theory}
World Scientific, 1994, 129}
\lref\nakheis{
H. Nakajima, ``Heisenberg algebra and
Hilbert Schemes of points on projective
surfaces,'' alg-geom/9507012}
\lref\sen{A. Sen, ``U-duality and Intersecting D-branes,''
hep-th/9511026}
\lref\stromvaf{A. Strominger and C. Vafa,
``Microscopic Origin of the Bekenstein-Hawking Entropy,''
hep-th/9601029}
\lref\orbcft{L. Dixon , D. Friedan, E. Martinec, S. Shenker,
``The Conformal Field Theory of Orbifolds,''
Nucl.Phys.B282:13-73,1987.}
\lref\bianchi{M. Bianchi, F. Fucito, G. Rossi, and
M. Martellini, ``Explicit construction of Yang-Mills
instantons on ALE spaces,''  hep-th/9601162}
\lref\vafadb{C. Vafa, ``Instantons on D-branes,''
hep-th/9511222;
``Gas of D-Branes and Hagedorn Density of BPS States,''
hep-th/9511088}

%
\def\quatthree{(7.1)}
\newsec{Introduction}

D-branes \refs{\dlp,\pol}\ are explicit realizations of
RR charged BPS states in superstring theory.

Witten \witten\ proposed that a $5$-brane in type \I\ string theory
is the zero size limit of the gauge $5$-brane solution of \chs,
built around a conventional gauge theory instanton.
Furthermore, the moduli space of instantons is realized as an
ADHM hyperkahler quotient.

ALE spaces are interesting because they describe the blowups of
K3 singularities, and because the metrics and Yang-Mills instantons
are explicitly computable.

We show that placing $5$-branes at an orbifold fixed point produces an
effective field theory whose vacua are points in instanton moduli space
on the {\it resolved} ALE space.
As in Witten's work, the $N=1$ supersymmetry of the $d=6$
D-brane world-volume theory leads to a hyperkahler quotient description
of the space of vacua.
A new element of the construction is a direct identification between
the NS-NS gravitational moduli and Fayet-Iliopoulos terms in the world-volume
theory, which provides a very simple way for moduli which blow up the
orbifold to couple to the world-volume theory.
The results justify a rather surprising claim:
by adding these couplings, we get an exact description of D-branes
moving on the resolved ALE space.

The simplest case is $U(N)$ instanton moduli space,
for which an ADHM construction has been developed by
Kronheimer and Nakajima. \kn\ %
This construction describes both instanton moduli spaces and the actual metric
on the ALE space.

To get this we want to start with type \II\ theory,
but as is well known the theory with $U(N)$ gauge group is anomalous.
A simple way around this is to work with a well-defined theory containing
$p$ and $p-4$ branes with $p<9$.
The resulting construction is identical to that of Kronheimer and Nakajima.

In a companion paper  \dmns\  the type I quivers defined in
this paper are used to construct  $SO(w)$ and
$USp(w)$ instantons on ALE spaces.

\subsec{Overview}

The body of the paper is the explicit construction of the
world-volume Lagrangian for a set of D-branes located at the fixed point in the
orbifold $\BC^2/\BZ_n$, followed by a discussion and mathematical
interpretation of the space of vacua.
Section 2 reviews the closed string spectrum and gravitational moduli,
and properties of the smooth ALE produced by blowing up the
orbifold singularity.

The D-brane world-volume theory will be a supersymmetric gauge theory --
for $5$-branes, a $d=6$, $\CN=1$ theory, and for $p<5$ essentially its
dimensional reduction (but not precisely -- see section 6).
Its spectrum is derived by imposing the point group and (for type \I) twist
projections on $U(N)$ gauge theory.
Much of the work here is a careful analysis of the consistency
conditions on the combined point/twist group (in section 3) and
its representations (in sections 4 and 5).
The results are easy to state by using quiver diagrams
(introduced in subsection 4.2),
and are given in figures 1-11 in sections 4 and 5.

The gauge Lagrangian is augmented by various couplings to the closed string
bulk and twisted fields, derived in sections 6, 7 and the appendix.
In section 7 we show that twisted moduli couple in the world-volume
Lagrangian as Fayet-Iliopoulos terms, using the argument of \dsw:
they are supersymmetry partners of a scalar required for $U(1)$ anomaly
cancellation, and also by world-sheet computation.
Combining this with the quiver diagrams, we have the complete D-flatness
conditions
which determine the space of vacua.
As is well known, these conditions are a physical realization of
the hyperk\"ahler quotient construction (subsections 6.1 and 6.5).

We proceed to compare these results with the work of Kronheimer and
Kronheimer and Nakajima in sections 8 and 9, and show that
these theories, derived by working in the
orbifold limit, in fact describe a finite region in moduli space.
In section 8 we show this for the ALE space itself, in a theory
containing a single D-brane (and its orbifold images), and in section 9
for the moduli space of $U(N)$ instantons on the ALE space.

Section 10 contains conclusions.

\newsec{Closed strings on  ALE spaces}

\subsec{ ALE spaces and orbifolds of $\BC^2$  }

Here we briefly review a few properties of ALE spaces
and define notation. For more information
see
\ref\egh{Eguchi, Gilkey, and Hanson, ``Gravitation,
Gauge Theories, and
Differential Geometry,''  Phys. Rep. {\bf 66}(1980)214}
\hitchiniii \kronheimer\hitchini.
An ALE space or gravitational instanton $\CM_\Gamma$ is a $4$-manifold
with anti-self-dual (\hk) metric asymptotic to $R^4/\Gamma$, where $\Gamma\in
SU(2)$
is a discrete subgroup.
When $\Gamma=\IZ_n$ an explicit description
of the gravitational instanton $X_n$ is available in the form of  the
multi-center Eguchi-Hanson gravitational
instanton \egh
\eqn\i{
\eqalign{
ds^2 &= V^{-1}(d t+ \vec A \cdot d \vec x )^2 + V dx^2\cr
V & =   \sum_{i=1}^{n} {1\over \vert \vec x-\vec x_i\vert} \cr
- \vec \nabla V   & = \vec \nabla \times \vec A \cr}
}
Here $t$ is an angular coordinate, $\vec x, \vec x_i$
are points in $\IR^3$.
%
%
Euclidean motions on the  $n$ vectors $\vec x_i$ produce
equivalent metrics, while otherwise inequivalent
$\vec x_i$ produce inequivalent metrics. The
moduli space of  such instantons is therefore the
$3n-6$-dimensional (for $n>2$) configuration space of $n$ points
in $\IR^3$. In the limit $\vec x_i\rightarrow 0$, or,
equivalently $\vec x \rightarrow \infty$ the metric
\i\ is easily seen to degenerate to the metric on the
orbifold $\IC^2/\Gamma$.

The coordinates in  \i\   degenerate along
line segments between the $\vec x_i$.
In fact, for generic $\vec x_i$ the  manifold $X_n$
is smooth and has
nontrivial topology:
$\Gamma$ is associated with a simply-laced
Dynkin diagram $D_\Gamma$ with $r_\Gamma$ nodes and
Cartan matrix $C_\Gamma$ in a well-known way and
this appears in the homology:
$H_2(\CM_\Gamma,\BZ)\cong \BZ^{r_\Gamma}$,  and the
intersection form   $-C_\Gamma$ identifies it with
the root lattice of $D_\Gamma$.
\foot{We will denote the
  extended Dynkin diagram and Cartan matrix
by $\tilde D_\Gamma$
and  $\tilde C_\Gamma$ respectively.}
A choice of ordering
of the $\vec x_i$ corresponds to a choice of
simple roots. The cohomology group
$H^2(\CM_\Gamma,\BZ)$ is identified with the weight lattice
and is spanned by a basis of anti-selfdual normalizable
two-forms. The  three covariantly constant self-dual
symplectic
forms $\vec \omega$  are not normalizable.

For $\Gamma=\IZ_n$ we may  choose a basis
$\Sigma_i $ for $H_2(\CM_\Gamma,\BZ)$
corresponding to $\vec x_i \vec x_{i+1}$.
The periods of the three
  symplectic forms $\vec \omega$ are
\hitchiniii:
\eqn\periods{
\int_{\Sigma_i} \vec \omega = \vec x_{i+1} - \vec x_i \equiv \vec \zeta_i
}
It is often convenient to let the indices take values
modulo $n$.
We also often write formulae with respect to a choice of
complex structure. Then the three symplectic
forms become the Kahler form $\omega^R$ and the
holomorphic  $(2,0)$ form $\omega^C$.
When we wish to emphasize the dependence of
the manifold on the gravitational moduli we
will write $\CM_\Gamma = X_n(\vec \zeta)$.  The ``global  Torelli
theorem,''  \kronheimer\  asserts that the
periods and asymptotic behavior determine
the metric uniquely.
Moreover, if $\psi$ is any automorphism of the root lattice
then $X(\psi(\vec \zeta)) \cong X(\vec \zeta)$. In particular
$X(-\vec \zeta) \cong X(\vec \zeta)$.
 Finally,  if $\vec \zeta\cdot \alpha=0$ for a root $\alpha$ then
$X(\vec \zeta)$ is singular since the 2-cycle associated
to $\alpha$ has   zero volume.

There is a third point of view on ALE spaces.
We may regard $\IC^2/\IZ_n$ as the affine algebraic
variety $X^n + YZ=0$ in $\IC^3$. The singularity
at the origin has a smooth resolution by
an algebraic variety  $\widetilde{\IC^2/\IZ_n}$.
 From this point of view the
nontrivial spheres $\Sigma_i$ constitute the
exceptional divisor of the blow-up.
 When $\zeta^C=0$ the ALE space is biholomorphic to
$\widetilde{\IC^2/\IZ_n}$, otherwise  is just diffeomorphic.
This last point of view makes contact with the
physical picture of resolving an orbifold singularity
by turning on blowup modes.

\subsec{Sigma model on ALE}

 Let $(z^1,z^2)$ be complex coordinates on $\BC^2$, with world-sheet
supersymmetry partners $(\po,\pt)$ (left moving) and
$(\tpo,\tpt)$ (right moving).
We let $\Gamma=\BZ_n$ act with fixed point $z=0$,
as
\eqn\sphol{
g(z^1,z^2) = (\xi \zo, \xi^{-1} \zt) .
}
where
$\xi=\exp 2\pi i/n$.
It will be useful to exhibit the original rotational symmetry $SO(4)$
as $SU(2)_L\times SU(2)_R$ by writing
\eqn\quatone{\QZ \equiv \left(\matrix{\zo&-\zbt\cr \zt&\zbo}\right)
= \QZ^{A'A};\qquad\qquad
\QZ \rightarrow g_L \QZ g_R.}
and embedding the twist in $SU(2)_L$.
Then $\vec \omega = -{1 \over  4} \tr \vec \sigma d \BZ^\dagger d \BZ$.
The unbroken $SU(2)_R$ acts on the sphere of complex structures of
this hyperk\"ahler manifold.
The sigma model on ALE target has $(4,4)$
supersymmetry, and contains $SU(2)_{k=1}$ current algebras
on both left and right;
\eqn\rsym{\Psi \equiv \left(\matrix{\po&-\pbt\cr \pt&\pbo}\right);
\qquad\qquad
\vec J=\tr\Psi\vec\sigma\Psi^{+}.
}
The orbifold
sigma model may be perturbed by exactly marginal fields in
the twisted sectors to obtain a family of sigma models
with target $X_n(\vec \zeta)$. The $N=(4,4)$ supersymmetry
survives and the somewhat special holonomy
\sphol\ in $SU(2)_L$ becomes generic.

When the sigma model is used as part of a ``compactification''
of a string theory  then, since
$\CM_\Gamma$ has $SU(2)$ holonomy the transverse
  $d=6$ field theories for  type \I, \IIa\ and \IIb\ strings on
$\BR^6\times \CM_\Gamma$ have $(0,1)$, $(1,1)$ and $(0,2)$ supersymmetry
respectively.
\foot{We are not really discussing compactification since the
ALE space is noncompact. Thus, the 6d theory will have a
continuous spectrum of particles. We will concentrate on
the modes which would be part of a massless spectrum if the
ALE space were compactified. For example, one may
imagine that the ALE space serves as a local description
of a singularity in a K3 manifold. }
The unbroken $SU(2)$  becomes
the $SU(2)_R$ of $d=6$ supersymmetry.
The left and right current algebras
\rsym\
 do not lead to symmetries of the string theory, but
only of the low energy limit (and at leading order in $\lambda$ and $\ap$).
They produce independent left and right $SU(2)_R$'s in type \IIa,
while in \IIb\ they sit in a $USp(4)$ not manifest on the world-sheet.
In type \I\ theory, the two $d=6$ supersymmetries are related as
$\tilde\epsilon=\epsilon$, and the left and right $SU(2)_R$'s
are also related, leaving their diagonal subgroup unbroken.
We will now list the massless closed string spectrum for the
three theories under consideration in terms of their
quantum numbers under
$$[SU(2) \times SU(2)]_{\rm little group} \times SU(2)_R^{diag}, $$
their Kaluza-Klein origin, and their orbifold realization.
Although this is standard and straightforward it will be
useful to have a summary of these states.

\subsec{Massless spectrum: \IIa}

We assume the  $SU(2)_L$ holonomy on
the ALE space space is generic. Under the
decomposition of transverse Lorentz
groups $[SU(2) \times SU(2) ] \times SU(2)_L\times SU(2)_R\subset
SO(8)$ we have the decompositions:
\eqn\spnrs{
\eqalign{
8_v & = (2,2;1,1) + (1,1;2,2) \cr
8_s & = (2,1;2,1) + (1,2;1,2) \cr
8_c & = (2,1;1,2) + (1,2;2,1) \cr}
}
$\CN=(1,1)$ representations are most conveniently
summarized by the $[SU(2) \times SU(2)]_{\rm little group} \times SU(2)_R$
content of the bosonic fields. The untwisted sector contains
the  $(1,1)$ gravity multiplet\nl
\def\spacing{\hbox{\qquad\qquad} }
\spacing NS-NS:  $(3,3;1) + (3,1;1) + (1,3;1) + (1,1;1)$\nl
\spacing R-R: $ (2,2;1)+  (2,2;3) $\nl
with a $(1,1)$ matter multiplet:\nl
\spacing NS-NS:  $(1,1;1) +  (1,1;3)$\nl
\spacing R-R: $ (2,2;1)$\nl
The   triplet  of scalars $(1,1;3)$ is obtained from the
KK reduction of $B$ along the three SD symplectic
   forms $\vec \omega$.

In addition there are $(n-1)$ $\CN=(1,1)$ matter
multiplets   associated to two-cycles $\Sigma_k$ of
\periods.  In the NS sector  the   state $(1,1;1)$ is obtained
by KK reduction
%
$  b^{(0)}_k   = \int_{\Sigma_k} B $. The
triplet $(1,1;3)$ states
are associated to the independent complex structure
and K\"ahler deformations which change $\vec\zeta_k $.
We denote the associated scalar fields by $\vec{\phi}_k$.
The RR vectors come from KK reduction:
${}^{6}C^{(1)}_k = \int _{\Sigma_k} {}^{10}C^{(3)} $.
Due to the many occurances of RR differential forms
in various dimensions we have adopted the
notation ${}^dC^{(q)}$ to denote a $q$-form field  in
$d$-dimensions.

%
%
%

When $\vec \zeta \rightarrow 0$ $X_n(\vec \zeta)$ reduces to
an orbifold and one can write the vertex operators for
the above states explicitly.
Of particular interest are the fields $\vec\phi_k$ which will come
from the NS-NS twisted sectors.

We denote states and fields in the sector twisted by $z_1(2\pi)=\xi^j z_1(0)$
as (for example) $\vec{\tilde\phi_j}$.  It will turn out (in section 8) that
this twisted sector basis is Fourier dual to the basis of two-cycles.
Since $\det g=1$ the lowest dimension NS-NS twist field in each sector
has $(h,\bar h)=(\half,\half)$.
Taking the twist to act as \quatone\ on both $\psi^i$ and $\tilde\psi^i$
gives us the massless fields
\def\mm{{-1/2+k/n}}
\eqn\twisty{
\eqalign{
\tilde \phi_k^{AB}(p)
&
\left(\matrix{\bar\psi^{\bar 1}_\mm\cr -\psi^2_\mm}\right)^{A}\otimes
\left(\matrix{\tilde\psi^1_\mm\cr
\tilde{\bar\psi}^{\bar 2}_\mm}\right)^{B}\ \ket{p;k;NS,NS},
1\leq k<n/2\cr
\tilde \phi_k^{AB}(p)
&
\left(\matrix{\psi^{ 1}_{-\half+(n-k)/n} \cr
-\bar\psi^{\bar 2}_{-\half+(n-k)/n} }\right)^{A}\otimes
\left(\matrix{\tilde{\bar\psi}^{\bar 1}_{-\half+(n-k)/n}\cr
\tilde{\psi}^{2}_{-\half+(n-k)/n}}\right)^{B}
\kern-10pt\ket{p;k;NS,NS},
 n/2\leq k\leq n\cr}
}
Here $\psi$'s are worldsheet fermions, on which tilde denotes right-mover.
The spacetime fields $\tilde \phi_k^{AB}$ are complex
  fields satisfying the reality condition
\eqn\IIrealtw{\tilde \phi_k^{AB} = \epsilon^{AC} \epsilon^{BD}
(\tilde \phi_{n-k}^{CD})^* \quad .}
The result for $k=n/2$ is obtained by quantizing the
Clifford algebra of zero modes, choosing a ground state
and applying the GSO projection.
Choosing the ground state to be annihilated by the imaginary
parts of all fermions in \twisty, it will be the limit $k\rightarrow n/2$
of \twisty, again satisfying \IIrealtw.
Together with the twisted RR sectors we get the bosons of
$n-1$ matter multiplets as described above.

Since the $SU(2)_L$ holonomy is nongeneric for
$\vec \zeta=0$ there will be additional
 matter multiplets in the $3$ of
$SU(2)_L$, which can   massless in the orbifold limit.
This produces an extra $3$ multiplets for
$\BZ_2$ and $1$ extra multiplet for $\BZ_n$, $n>2$.
Note that these are `bulk' modes and thus non-normalizable on $\CM_\Gamma$.

\subsec{Massless spectrum: \IIb}

Repeating the above discussion for the \IIb\ string
we have the $\CN=(2,0)$ gravity multiplet:\nl
\spacing NS-NS:   $(3,3;1) + (1,3;1)$\nl
\spacing R-R:  $(1,3;1) + (1,3;3) $\nl
In the untwisted sector there are
  two matter multiplets. The first, containing
the self-dual projection $B^+$ of $B_{\mu\nu}$
  and the dilaton is:\nl
\spacing NS-NS:   $(3,1;1) + (1,1;1)$\nl
\spacing R-R:  $(1,1;1) + (1,1;3) $\nl
The $(1,1;3)$ RR states come from KK reduction of
the two-form ${}^{10}C^{(2)}(x,y) = {}^{6}C^{(0)}_a(y) \omega^a(x) +\cdots $
along $\vec \omega$.

The second matter multiplet  containing the internal
volume and the reduction of $B$ along $\vec \omega$ is \nl
\spacing NS-NS:   $(1,1;1) + (1,1;3)$\nl
\spacing R-R:  $(3,1;1) + (1,1;1) $\nl
As in the \IIa\ theory there
  are $(n-1)$ matter multiplets associated to
the 2-cycles $\Sigma_k$.
The   NS-NS states $(b_k^{(0)}, \vec{\tilde \phi}_k)$
are obtained exactly
as in the \IIa\ case. The  RR fields in
$(3,1;1) + (1,1;1) $  are obtained from projection
of the 10d RR forms along $\Sigma_k$:
\eqn\projrr{
\int_{\Sigma_k} {}^{10}C^{(2)} = {}^{6}C^{(0)}_k
\qquad
\int_{\Sigma_k} {}^{10}C^{(4)} = {}^{6}C^{(2)}_k
}
In the orbifold limit the NS-NS states are obtained
exactly as in \twisty.

%
%
%
%

\subsec{Massless spectrum: I}

Making an orientation projection on the \IIb string
gives the massless closed string
 sector of the type I string.
The untwisted sector gives a $(1,0)$ gravity
multiplet, a tensor multiplet ($(3,1;1)+(1,1;1)$),
a hypermultiplet, and additional hypermultiplets on orbifolds.

Applying the $\Omega$ projection to the states \twisty\ in
a twisted sector gives a (linear) hypermultiplet:
 Using \rsym\ and \twisty, we see that
the  NS-NS scalars form a $(1,1;3)$.
These are the metric moduli which change $\vec \zeta_k$.
The fourth scalar in the $(1,1;1)$ is the
  RR state  $ {}^{6}C^{(0)}_k = \int_{\Sigma_k}  {}^{10}C^{(2)} $.

\newsec{Adding Dirichlet $5$-branes.}

We define D-branes on orbifolds of $\BC^2$ by
first defining D-brane configurations on $\BC^2\times \BR^6$
and then extending the action of the orbifold point group to the open string
sectors.
If $x$ is an allowed endpoint for open strings, all of its
images under the point group must also be allowed endpoints -- thus
each D-brane will be represented by the set of its images under the point
group.
Such a formalism has recently been discussed
by Gimon and Polchinski \gimon, and we review and add to their results here.

A D$p$-brane relates the two supersymmetries of type \II\ theory
as $\tilde\epsilon=\Gamma_D\epsilon$, where
$\Gamma_D=\epsilon_{\mu_1\ldots\mu_{p+1}}
 \Gamma^{\mu_1}\ldots\Gamma^{\mu_{p+1}}$
and $\epsilon_{\mu_1\ldots\mu_{p+1}}$
is the $p+1$-dimensional volume form $\epsilon_{\mu_1\ldots\mu_{p+1}}
	dX^{\mu_1}\wedge\ldots\wedge dX^{\mu_{p+1}}$.
Thus the maximal supersymmetry in the world-volume
theory after orbifold projection will be $N=1$ in $d=6$.

One can add several D-branes and preserve this supersymmetry
if the conditions $\tilde\epsilon=\Gamma_D\epsilon$ are compatible.
The theories we consider of parallel $p$-branes contained within
$p+4$-branes are such a case.

We first treat the subsector of $N$ $5$-branes, each filling $\BR^6$
and located at a point $x$ in $\BC^2$.
Each open string sector is labelled by a Chan-Paton index $i$ at each end.
Let $\CS$ be the set of these $N$ indices and $V\equiv \BC^N$.
In type \II\ theory each index $i\in\CS$
will label a single D-brane, whose position will be $x(i)$.
Let $A_\mu(x)$ be a $n\times n$ hermitian matrix gauge field
related to an open string state as
\eqn\state{\ket{A} = A_\mu(x)^{ij} \psi^\mu \ket{0_{NS};i,j}.}
For $p<9$, let $X_I$ be the
$N\times N$ matrix of scalars produced by dimensional reduction.

To define the orbifold, we must define an action $\gamma$
of the point group $G_1$ on $\CS$,
correlated with the positions of the D-branes
in space: $g(x(i)) = x(\gamma(g) (i))$.
The resulting theory will be a truncation of the original
super Yang-Mills theory and we will describe this in terms of the
action of the point group on the gauge fields $A$ and scalars $X$ of
this theory.  The action must preserve the
inner product $\tr A^+ B$, and the string joining interaction $AB$.
Thus $\gamma(g) $ must act as
\eqn\stract{g:A_\mu(x) \rightarrow \gamma(g) \ A_\mu(x')\ \gamma(g)^{-1}}
with $\gamma(g)$ unitary.
The action on fields with a vector index in $\BC^2$
also includes a rotation on the space indices,
\eqn\orbrot{g: X^I(x) \rightarrow  R(g)^I_J~~ \gamma(g)  X^J(x') \gamma(g)
^{-1}.}
Fields surviving the orbifold projection are invariant under
the action \stract\orbrot\ and hence the
  unbroken gauge symmetry will be the commutant of this
representation in $U(N)$.

We now extend the Chan-Paton indices $i\in\CS$
to label the entire set of D-branes.
In all cases, the action \stract\ applies for fields with indices transverse
to the orbifold, while \orbrot\ applies to fields with vector indices in
the orbifold.

The theory now contains ``DN'' open string sectors with one end on a
$p+4$-brane and the other on a $p $-brane.  For $p=5$ these will produce
massless hypermultiplets, whose scalars transform
in the doublet of $SU(2)_R$.
As discussed in \witten\douglas\joerev\
the fields carry an $SU(2)_R$ index $A$ from
quantization of the zeromodes of $\psi^{6,7,8,9}$ in
addition to a $p+4$-brane index $M$ and a $p$-brane
index $m$.  This gives scalar fields $h^{Am}_{~~M}$
for strings oriented from the $p+4$ to the $p$-brane and
$\tilde h^{AM}_{~~m}$ for strings oriented the
other way.  The two orientations are related
by a reality condition:
 \eqn\hyperreality{\epsilon^{AB} \bigl(h^{Bm}_{~~M}(x)\bigr)^*
= \tilde h^{AM}_{~~m} (x)  \quad . }
  The point group does not act on $SU(2)_R$, so invariant
DN fields satisfy:
\eqn\orbhyp{h^A(x) =  \gamma(g)  h^A(x') \gamma(g) ^{-1}.}

Defining a type \I\ theory requires introducing $9$-branes,
and giving the action of the orientation reversal $\Omega$.
This acts on the fields of a general $p$-brane theory as
\eqn\om{\eqalign{
A_\mu(x) &= -\gamma(\Omega)  A_\mu^{tr}(x) \gamma(\Omega) ^{-1} \cr
X^I(x) &=  \gamma(\Omega)  X^{I,tr}(x) \gamma(\Omega) ^{-1}\cr
\epsilon^{AB} \bigl(h^{Bm}_{~~M}(x)\bigr)^*
&= (\tilde h^A)^{tr}(x) =
\alpha i   (\gamma(\Omega) )_{mm'} h^{Am'}_{~~M'} (x) (\gamma(\Omega)
^{-1})^{M'M} .\cr
}}
where $\alpha=\pm 1$.
The relative minus sign between $A$ and $X$
is determined by standard world-sheet considerations,
while the $\pm i$ in the action on $h$ was explained by Gimon and Polchinski
\gimon.
$\gamma(\Omega) $ must also be unitary. We may
absorb $\alpha$ into the definition of $\gamma(\Omega)_{mm'}$.

\subsec{A remark on consistency conditions}

We now examine the consistency conditions on
the matrices $\gamma(g) , \gamma(\Omega) $. We will
restrict attention to algebraic consistency conditions,
and not consider consistency conditions following from
tadpole cancellation.~\gimon\ %
Such conditions are generally of the form $0 = \int dH = \sum ({\rm sources})$
where the integral is zero on a compact space,
and one justification for this neglect
is that we are working with a non-compact space.  Configurations which do
not cancel the tadpole are sensible configurations with non-zero charge.

This is not completely satisfactory as there are configurations which cannot
be interpreted this way.  For example, the original type \I\ anomaly
cancellation which required $SO(32)$ (on $\BR^{10}$)
is phrased as a cancellation between
$9$-brane and non-orientable closed string tadpoles.
These produce a zero-form
on the right which is not a source of a physical field.  One might also be
interested in studying instantons on compact spaces.

To deal with these situations, one can lower the dimensions of the D-branes,
so that they occupy a subspace of $\BR^6$, and can serve as physical sources.
The resulting world-volume theories are essentially dimensional reductions of
$5$ and $9$-brane theories, with the same supersymmetry.
In type \II\ this is easy, while in type \I\ to make complete sense of this one
must consider an orientifold of $\BR^6$ not containing the $\Omega$ projection,
and preserving some supersymmetry, such as the T-dual of type \I\ \dlp\ in
four of the six dimensions.  One is free to take the D-branes away from the
new fixed points.

The conclusion is that for the purpose of studying moduli spaces of D-brane
configurations (in up to six dimensions), the tadpoles can be ignored.

\subsec{Algebraic consistency conditions}

We begin with some general remarks. Consider a
string theory in the soliton sector where D-branes
are wrapping various supersymmetric cycles
$\CB_i$.
The one-string Hilbert space  includes sectors
associated to pairs of wrapped cycles:
 $\hat \CH_{\CB, \CB'} $.  These are
called ``DD sectors'' for $\CB= \CB'$ and
``DN sectors'' otherwise.  If $\CB$ has $n$-wrapped
D-branes then we associate a vector space
$V_\CB=\IC^n$ to the cycle and the  Hilbert space is of the form:
\eqn\prodhil{
\hat \CH_{\CB, \CB'}
= \CH_{\CB, \CB'}\otimes End(V_\CB, V_\CB')
}
for strings oriented from $\CB$ to $\CB'$. Here
$\CH_{\CB, \CB'}$ is a chiral conformal field theory
specified by boundary conditions (see for example,
\ref\cardy{J. Cardy, ``Boundary Conditions, Fusion Rules and the Verlinde
Formula,'' Nucl.Phys.B324:581,1989} )
 while $End(V_\CB, V_\CB')$, the space of
all linear transformations, are just the Chan-Paton factors.
\foot{It is argued in \gimon\  that choosing a subspace
leads to inconsistent  dynamics.}

The action of the orbifold group $G_{orb}$ on $\hat \CH$ is
 defined by
\eqn\abstrac{
\hat U(g)(\phi\otimes \lambda)
= (U_{\CB\CB'} (g) \cdot \phi) \otimes \gamma_\CB(g) \lambda  \gamma_{\CB'}
(g)^{-1}
}
In the IIB string we can make a further projection by the
orientation operator $\Omega$. The full orbifold group
$G^{tot}$ defining the type I theory on an orbifold
is then a $\IZ_2$  extension of $G_{orb}$:
$$
1 \rightarrow \IZ_2 \rightarrow G^{tot} \rightarrow
G_{orb} \rightarrow 1
$$
In this paper,
we will make the minimal assumption that the
extension is trivial $\IZ_2 \times G_{orb}$. The action of
$\Omega$ then takes the form:
\foot{Note that our conventions for the action on the Chan-Paton
factors differ slightly from \gimon.}
\eqn\abstracii{
\hat U(\Omega)(\phi\otimes \lambda)
= (U_{\CB\CB'}(\Omega) \cdot \phi)
\otimes \gamma_\CB(\Omega) \lambda^{tr} \gamma_{\CB'}(\Omega)^{-1}
}

The algebraic consistency conditions state that the
representation $\hat U$ must be anomaly-free, that is,
it must give a true (not projective)  representation of the orbifold
group $\IZ_2 \times G_{orb}$. Let us consider first
the DD sectors. The action of the group on the
CP factors is adjoint so the representation on
$\CH_{\CB\CB}$ must be
anomaly free.  By Schur's lemma
$\gamma_\CB$ must satisfy the relations of
the group up to scalar factors.
Specializing  to the   case $\IZ_2 \times \IZ_n$
we find the conditions:
 \eqn\actcond{
\eqalign{
\Omega^2 =1: \qquad
\gamma_\CB(\Omega)
&
= \chi_\CB(\Omega )  \gamma_\CB(\Omega)^{tr}
\cr
 \Omega g = g \Omega: \qquad
\gamma_\CB(g) \gamma_\CB(\Omega) \gamma_\CB(g)^{tr}
&
= \chi_\CB(g,\Omega)  \gamma_\CB(\Omega)
\cr
g^n=1: \qquad
\gamma_\CB(g)^n & = \chi_\CB(g ) 1 \cr}
}
where $\chi_\CB(\Omega), \chi_\CB(g,\Omega), \chi_\CB(g)$
are scalars.
The choice of $\gamma$-matrices is of course not
unique. First,
a unitary change of basis on the Chan-Paton
spaces $V_\CB$ acts by
\eqn\chgbs{
\eqalign{
\gamma(g) & \rightarrow U\gamma(g) U^{-1} \cr
\gamma(\Omega) & \rightarrow U\gamma(\Omega)  U^{tr}\cr}
}
Second,  we may redefine the matrices
$\gamma$ by a scalar factor
$\gamma_\CB \rightarrow
\epsilon_\CB(g) \gamma_\CB(g)$
etc.

Now let us consider the consistency
conditions on the   $\chi$-factors.
Since $\gamma$ are unitary matrices,
all such factors  are phases.
Consistency requires
$\chi_\CB(\Omega)=\pm 1$ and
that $\chi_\CB(g,\Omega)$ is an
$n^{th}$ root of 1. By rescaling
$\gamma(g) $ we can set $\chi_\CB(g)=1$.
This still leaves the freedom of rescaling
$\gamma(g)$ by an $n^{th}$ root of
unity which changes
$\chi_\CB(g,\Omega)\rightarrow
\xi^2 \chi_\CB(g,\Omega)$.
Thus we can set:
\eqn\stwoposs{\chi_\CB(g,\Omega)=
\cases{1&(n\ {\rm odd})\cr 1, \xi&(n\ {\rm even}).}}
Different choices in \stwoposs\
lead to different physics.

There are further consistency conditions on the
$\chi$'s following from considerations of the
sectors $\hat \CH_{\CB\CB'}$ with
$\CB \not= \CB'$, i.e., the ``DN sectors.''
In these sectors two
 interesting new subtleties can occur.
First it can happen that  it is not the group
$ \IZ_2\times G_{orb}$ but
actually
a nontrivial extension $\CG$ of the orbifold
group:
$$
1 \rightarrow K \rightarrow\CG
\rightarrow  \IZ_2\times G_{orb}\rightarrow 1
$$
which
acts separately on the conformal field theory
and Chan-Paton factors in such a way that
$K$ acts trivially on the product $\hat \CH$.
Second, the group $\IZ_2\times G_{orb} $
(or an extension of it) can have a projective
(=anomalous) action on the separate factors
as long as  the combined representation $\hat U$ is
nonanomalous.

An example of the first subtlety has been
discussed by Gimon and Polchinski
\gimon. In $(p+4,p)$ sectors the
 $\IZ_2$ orientation group generated by
$\Omega$ is extended to $\IZ_4$:
$$
1 \rightarrow \IZ_2 \rightarrow
\IZ_4 \rightarrow \IZ_2
\rightarrow 1
$$
when acting on the CFT and Chan-Paton
DN factors separately.  Indeed,
\gimon\ showed that locality of the
operator product expansion implies that
$\Omega^2$ acts by $-1$ on the CFT space
  $\CH_{p+4,p}$.
The requirement that
the group $K$, generated by $\Omega^2$,
act trivially on $\hat \CH_{p+4,p}$ then
implies:
\foot{If we add the condition of tadpole
cancellation then in addition $\chi_9(\Omega)=+1$.}
\eqn\relchis{
\chi_{p+4}(\Omega) = - \chi_p(\Omega)
}
As mentioned above, this is the source of the factor
$\alpha i$ in \om.

The second subtlety entails the existence
of a group cocycle
$\epsilon_{\CB\CB'}\in
H^2(\IZ_2\times G_{orb},\IC^*)$
in the action on the CFT factor.
Cancellation of anomalies then
requires
\eqn\cnclcoc{
\epsilon_{\CB\CB'}(g,\Omega)
\epsilon_{\CB\CB'}( \Omega, g)
= \chi_\CB(g,\Omega) \chi_{\CB'}^{-1}(g,\Omega)
}
In this paper we will restrict attention to the
simplest case
\eqn\chiseal{
 \chi_\CB(g,\Omega)=  \chi_{\CB'} (g,\Omega)\qquad .
}
It would be very interesting to see if the
more general possibility \cnclcoc\  defines
consistent string theories. These would be new
discrete parameters needed to specify backgrounds,
analogous to
\ref\dt{C. Vafa, ``Modular Invariance and Discrete Torsion on Orbifolds,''
Nucl. Phys. B273:592, 1986. }
\ref\am{P. Aspinwall and D. Morrison, ``Chiral Rings Do Not Suffice: N=(2,2)
Theories with Nonzero Fundamental Group,'' hep-th/9406032}
\ref\apt{P. Aspinwall, ``An N=2 Dual Pair and a Phase Transition,''
hep-th/9510142}.

\newsec{Quiver Diagrams and the
 DD spectrum  of  the $p$-brane  at the fixed point}

The general situation is best discussed at a point of maximal
symmetry: we locate a set of D-branes at the fixed point,
choose an action of the point group,
compute the massless spectrum, and then give a geometrical
interpretation to the resulting moduli. In this section we consider
only the DD sectors.

\subsec{Type II}

We first discuss a type \II\ theory and a subsector of $p$-branes
of definite $p$.
This sector is determined by a choice of unitary representation of $\BZ_n$,
$V^{(p)}$.
This will be a sum of one-dimensional irreps $R_i$ on
which the generator $g$ of $\BZ_n$ acts as $\xi^i$, so the
representation is determined by the vector of their multiplicities $v^{(p)}_i$,
\eqn\defR{V^{(p)} = \oplus_{i=0}^{n-1} v^{(p)}_i R_i=
\oplus_{i=0}^{n-1} V^{(p)}_i.}
with $v^{(p)}=\sum_i v^{(p)}_i$.
The gauge symmetry $U(v^{(p)})$ is broken to
\eqn\unbuni{G = \otimes_i\ U(v^{(p)}_i).}

We will use a bi-index notation $A_{i{\alpha_i};j{\beta_j}}$
(with $0\le i,j\le n-1$ and $1\le \alpha_i,\beta_i\le v_i$)
for a matrix in the adjoint of $U(v)$, and usually abbreviate this to
$A_{i{\alpha},j{\beta}}$.
The massless gauge fields then satisfy the projection
\eqn\orbgpro{
A_{i{\alpha},j{\beta}} = \xi^{i-j} A_{i{\alpha},j{\beta}}}
leaving $A_{i{\alpha},j{\beta}}$ with $i=j$.

The projection \orbrot\ on the hypermultiplets is just as easy to solve.
We assemble $X^I$ into two scalar components $X, \bar X$ diagonalizing
the action of $R$. Then:
\eqn\orbpro{\eqalign{
X_{i{\alpha},j{\beta}} &= \xi^{i-j+1} X_{i{\alpha},j{\beta}}\cr
\bar X_{i{\alpha},j{\beta}} &= \xi^{i-j-1}
 \bar X_{i{\alpha},j{\beta}}}}
so $X$ will be ``block off-diagonal,'' the  nonzero blocks
being $X_{i,i+1}, \bar X_{i+1,i}$:
\eqn\matrep{
\eqalign{
X & = \pmatrix{
0 &  X_{01} & 0 & 0 & \cdots \cr
0 &  0 & X_{12} & 0 & \cdots \cr
 &  \cdots  &  & & \cdots \cr
 &  \cdots  &  & & \cdots \cr
X_{n-1,0 } & 0  & & \cdots & 0  \cr}\cr
\bar X & = \pmatrix{
0 &  0 & \cdots & & \bar X_{0,n-1}  \cr
\bar X_{10} &  0 &0  & & \cdots \cr
 0 &  \bar X_{21} &0  & & \cdots \cr
 &  \cdots  &  & \cdots & \cr
0 & 0  & \cdots & & 0  \cr}\cr}
}
Moreover,
under the gauge group \unbuni\   these scalars transform in the
representations:
\eqn\hypreps{
\eqalign{
X_{i, i+1}  & \in \bar v_{i+1}  \otimes v_i  \cong \Hom(V_{i+1}, V_i) \cr
\bar X_{i+1,i}  & \in  \bar v_{i} \otimes v_{i+1} \cong \Hom( V_{i},V_{i+1})
\cr}
}
Together, these two matrices of scalars comprise a matrix of
hypermultiplets.

\ifig\quivfig{A type II quiver diagram for D-branes transverse to $X_n$,
$n=1,2,3,\dots$.
This figure represents the field content of the SYM theory on the transverse
3- or 4-brane. At  the
vertices we have vectormultiplets in the gauge group indicated, while on the
links we have hypermultiplets in representations determined
 by the fundamental representation at each vertex . Such a diagram with
oriented edges will be called a quiver diagram.}
{\epsfxsize3.0in\epsfbox{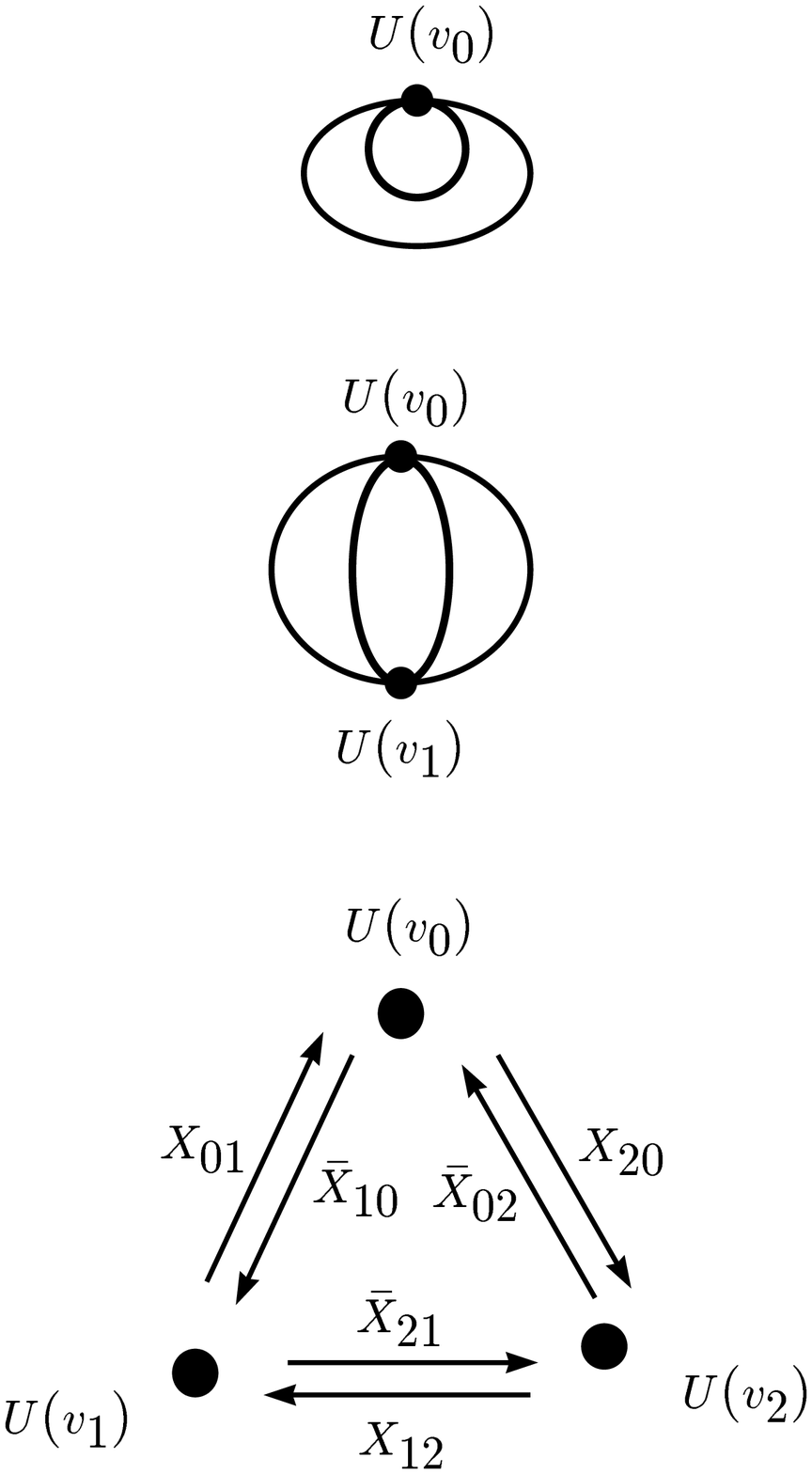}}

\subsec{Quiver diagrams}

The field content of the SYM theory on the p-brane may be
summarized using  a ``quiver diagram.''
In these diagrams we associate vector multiplets
with vertices and hypermultiplets with links.
A vertex will be associated with both a vector space $V$,
and the semisimple component of the gauge group which acts on $V$.
An oriented link from vertex $V_1$ to $V_2$ represents a complex
scalar  transforming in the representation
$\bar V_1 \otimes V_2\cong \Hom(V_1, V_2) $. Two
links with opposite orientation  comprise a single hypermultiplet.
Thus, for example, $(X_{i,i+1}, \bar X_{i+1,i})$ form hypermultiplets.
The field content is summarized in \quivfig.


It is worth remarking that, although this paper focuses on
the case $\Gamma=\IZ_n$, in fact most of the results
should generalize to arbitrary A-D-E ALE spaces.
These other spaces will be obtained from nonabelian
orbifolds.
In the other cases the diagram \quivfig\ will be
replaced by the extended Dynkin diagram
$\tilde D_\Gamma$.

\subsec{Canonical form for $\gamma(\Omega) $ }

%

The type \I\ effective theory can
be derived by further imposing the $\Omega$ projection.
We must find the most general solution of \actcond\ up to unitary
transformations.
We will work in the basis with $\gamma(g) $ diagonal, and
the last condition of \actcond\ then requires
\eqn\lastcon{(\gamma(\Omega) )_{i{\alpha},j{\beta}} =
 \chi(g,\Omega)\ \xi^{i+j}\ (\gamma(\Omega) )_{i{\alpha},j{\beta}}.}
Forcing $\gamma(\Omega) $ to have nonzero blocks
only for $\chi(g,\Omega)\ \xi^{i+j}=1$. We may still use
 the freedom to do transformations
\eqn\unittr{\gamma(\Omega)\rightarrow U\gamma(\Omega)\ U^{tr}}
with
$U \in \otimes_i\ U(v^{(p)}_i)$ to put $\gamma(\Omega) $ into
canonical form. The unbroken gauge group is then
determined from
\eqn\unbrgg{
U \gamma(\Omega) U^{tr} \gamma(\Omega)^{-1} = 1
}

We first consider the case $\chi(g,\Omega)=+1$.
The non-zero blocks are
\eqn\nzpblk{
(\gamma(\Omega) )_{i{\alpha};(n-i){\beta}} =
 \chi(\Omega) \ (\gamma(\Omega) )_{(n-i){\alpha};i{\beta}}^{tr}}
($i=n$ is identified with $i=0$).
For $i\ne 0$ and $i\ne n/2$ ($n$ even), the condition relates two
different blocks, and we require $v_i=v_{n-i}$.
Its general solution can be reduced to
\eqn\nzpsol{\eqalign{
(\gamma(\Omega) )_{i{\alpha};(n-i){\beta}} &= \delta_{\alpha,\beta} \qquad
0<i<n/2\cr
(\gamma(\Omega) )_{(n-i){\alpha};i{\beta}} &= \chi(\Omega)
\delta_{\alpha,\beta}
\qquad  n/2< i <  n \cr }}

For $i=0$ or $i=n/2$, the condition relates
$(\gamma(\Omega) )_{i{\alpha},i{\beta}}$ to its transpose.
By a transformation \unittr, this can be reduced to $\delta_{\alpha,\beta}$
if $\chi(\Omega)=+1$, while if $\chi(\Omega)=-1$, $v_0$ (or $v_{n/2}$) must be
even and
$\gamma(\Omega) $ can be reduced to the canonical skew-symmetric form
$\epsilon_{\alpha \beta}$.
\footnote*{To see this for $\chi(\Omega)=+1$,
write $\gamma(\Omega) =M+iN$ with $M$ and $N$ real.
Using $\gamma(\Omega) ^{-1}=\gamma(\Omega) ^+$ and the first line of \actcond\
one
can show that $[M,N]=0$ and are both symmetric, so can be simultaneously
diagonalized by \unittr\ with $g$ orthogonal.  Finally, \unittr\ with $g$
diagonal can be used to reduce the eigenvalues to $1$.
The argument for $\chi(\Omega)=-1$ is very similar.}

For $\chi(g,\Omega)=\xi$, \nzpblk\ is changed to
\eqn\nopblk{
(\gamma(\Omega) )_{i{\alpha},(n+1-i){\beta}} =
 \chi(\Omega)\ (\gamma(\Omega) )_{(n+1-i){\alpha},i{\beta}}^{tr}.}
For $n$ even, the blocks $i,n+1-i$ and $n+1-i,i$ related by these conditions
are always distinct.
Thus we require $v_i=v_{n+1-i}$. Moreover, the blocks can
always be diagonalized as in \nzpsol.

\subsec{Type I Quiver diagrams}

We now list the unbroken gauge groups for the effective
theory on a $p$-brane world-volume surviving after
the orbifold and orientation projections.
There are five cases to consider:

\ifig\quivfigi{ A type I quiver diagram for $X_n$, $n$ odd, $\chi(\Omega)=+1$.
Note that the hypermultiplet field content is determined by
$V_0 = \IC^{v_0}$. For $\chi(\Omega)=-1$ replace $O(v) \rightarrow USp(v)$.
Double arrows on the edges have been suppressed }
{\epsfxsize3.0in\epsfbox{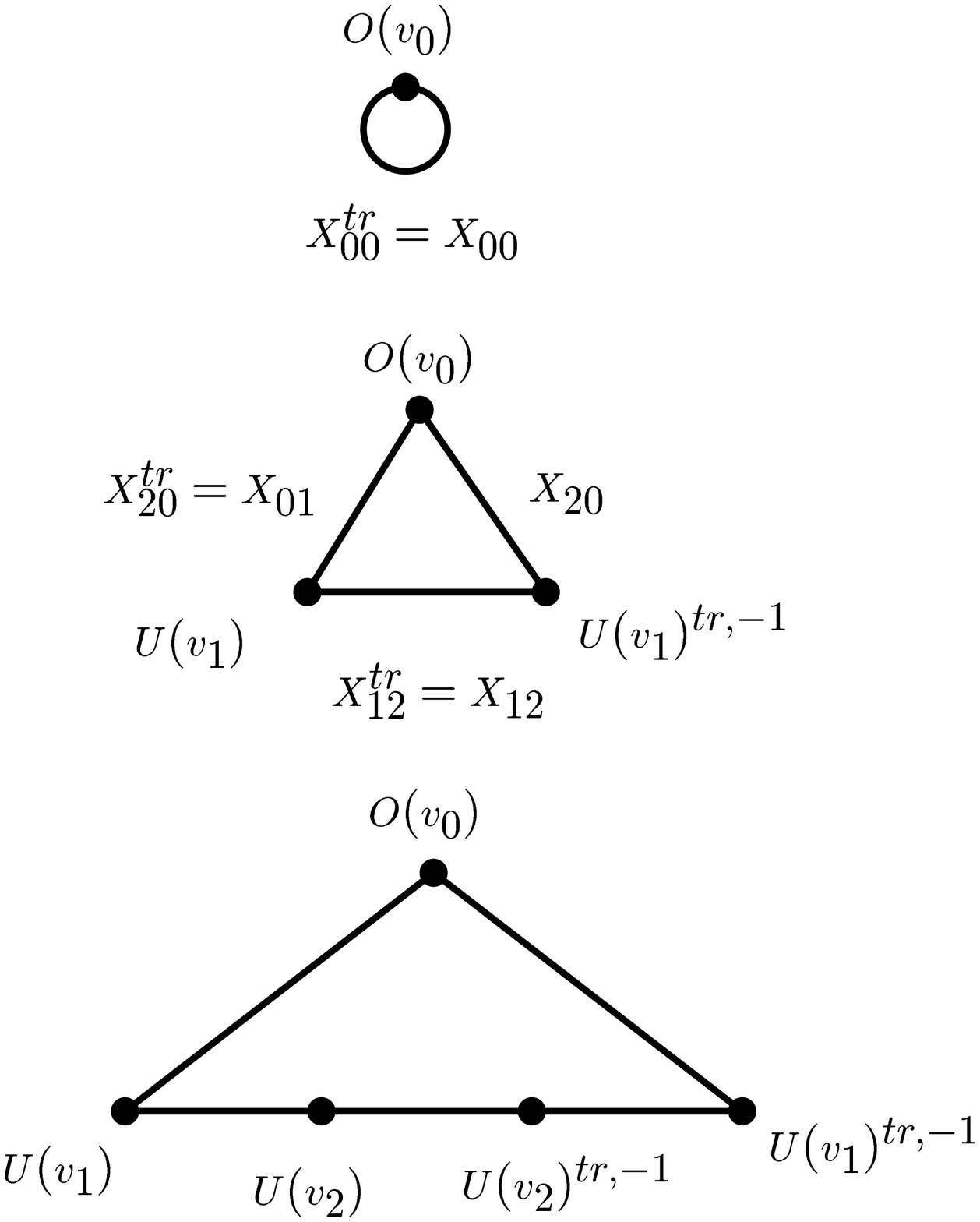}}

\item{I.1.} $\chi(\Omega)=+1, \chi(g,\Omega)=1$. $n$ odd.
\eqn\grpi{
\eqalign{
\gamma(\Omega) & =
\pmatrix{
1_{v_0} &    &     &    &    \cr
             &    &     &      &    1_{v_1} \cr
             &    &     &     1_{v_2}  &        \cr
             &    &     1_{v_2}  &       &        \cr
             &     1_{v_1  } &     &       &        \cr}  \qquad (n=5) \cr
G_1(\vec v) \equiv O(v_0)\times\bigl[
&
U(v_1) \times U(v_2) \times\cdots U(v_{(n-1)/2})\bigr] \cr
& = \{ (U_0, U_1, \dots, U_{n-1}) : U_i U_{n-i}^{tr} =1 \quad 1\leq i \leq n-1
\} \cr}
}
Moreover  $V_i = V_{n-i} $ and  the conditions on the hypermultiplets are:
\eqn\hyphomi{
\eqalign{
(X_{n-i-1,n-i})^{tr} & = X_{i,i+1}\in \Hom(\IC^{v_i}, \IC^{v_{i+1}} ) \cr
(\bar X_{n-i+1,n-i})^{tr} & = \bar X_{i,i-1}\cr}
}
The field content is summarized by the quiver diagram
\quivfigi. The above conditions may be interpreted as
saying that the diagram is symmetrical under reflection
about a vertical line through the vertex $V_0$.

\item{I.2.} $\chi(\Omega)=-1, \chi(g,\Omega)=1$. $n$ odd.  This is very similar
to
case I.1. We have a slightly different form for $\gamma(\Omega) $:
\eqn\grpii{
\eqalign{
\gamma(\Omega) & =
\pmatrix{
\epsilon_{v_0} &    &     &    &    \cr
             &    &     &      &    1_{v_1} \cr
             &    &     &     1_{v_2}  &        \cr
             &    &     -1_{v_2}  &       &        \cr
             &     -1_{v_1  } &     &       &        \cr}  \qquad (n=5) \cr
G_2(\vec v) \equiv USp(v_0)\times\bigl[
&
U(v_1) \times U(v_2) \times\cdots U(v_{(n-1)/2})\bigr] \cr
& = \{ (U_0, U_1, \dots, U_{n-1}) : U_i U_{n-i}^{tr} =1 \quad 1\leq i \leq n-1
\} \cr}
}
We again have  $V_i = V_{n-i} $, but the conditions on the
hypermultiplets become more complicated:
\eqn\hyphomii{
\eqalign{
X_{01}& = - (X_{n-1,0}\epsilon_{v_0})^{tr} \cr
X_{i,i+1} & = (X_{n-i-1,n-i})^{tr} \qquad 1\leq i \leq {n-3\over 2}\cr
X_{(n-1)/2, (n+1)/2} & = - (X_{(n-1)/2, (n+1)/2})^{tr}\cr
\bar X_{10}& =  (\epsilon_{v_0}\bar
X_{0,n-1})^{tr} \cr
\bar X_{i+1,i} & = (\bar X_{n-i,n-i-1})^{tr}
\qquad 1\leq i \leq {n-3\over 2}\cr
\bar X_{(n+1)/2, (n-1)/2} &
= - (\bar X_{(n+1)/2, (n-1)/2})^{tr}\cr}
}
Again we have a diagram analogous to \quivfigi\  with
reflection symmetry.

\ifig\quivfigii{ A type I quiver diagram for $X_n$, $n$ even,
$\chi(\Omega)=+1$.
For $\chi(\Omega)=-1$ replace $O(v) \rightarrow USp(v)$. }
{\epsfxsize3.0in\epsfbox{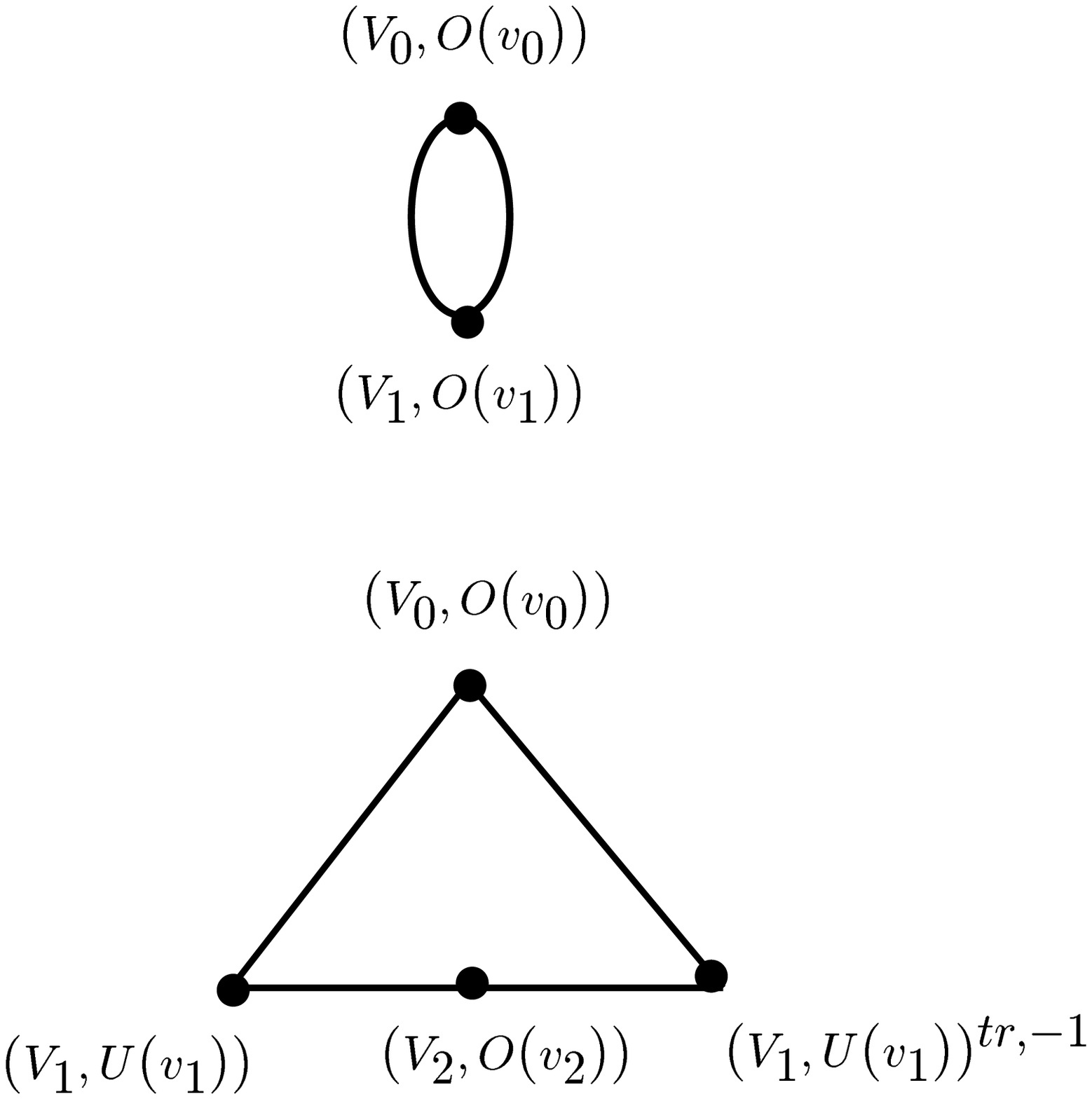}}

\item{I.3.} $\chi(\Omega)=+1, \chi(g,\Omega)=1$. $n$ even.
\eqn\grpiii{
\eqalign{
\gamma(\Omega) & =
\pmatrix{
1_{v_0}     &     &    &    \cr
             &    &           &    1_{v_1} \cr
             &    &          1_{v_2}  &        \cr
             &         1_{v_1} &         &        \cr}  \qquad (n=4) \cr
G_3(\vec v) \equiv O(v_0)\times\bigl[
&
U(v_1) \times U(v_2) \times\cdots    \times
U(v_{n/2-1})\bigr]
 \times O(v_{n/2})\cr
& = \{ (U_0, U_1, \dots, U_{n-1}) : U_i U_{n-i}^{tr} =1 \quad 0\leq i \leq n \}
\cr}
}
(We have simply $O(v_0)\otimes O(v_1)$ for $n=2$.)

The conditions on the scalar fields are \hyphomi.

\item{I.4.} $\chi(\Omega)=-1, \chi(g,\Omega)=1$. $n$ even.
\eqn\grpiv{
\eqalign{
\gamma(\Omega) & =
\pmatrix{
\epsilon_{v_0}     &     &    &    \cr
             &    &           &    1_{v_1} \cr
             &    &          \epsilon_{v_2}  &        \cr
             &         -1_{v_1} &         &        \cr}  \qquad (n=4) \cr
G_4(\vec v) \equiv USp(v_0)\times\bigl[
&
U(v_1) \times U(v_2) \times\cdots    \times
U(v_{n/2-1})\bigr]
 \times USp(v_{n/2})\cr
 = \{ (U_0, U_1,
&
\dots, U_{n-1}) : U_i U_{n-i}^{tr} =1 \quad 1\leq i \leq n-1, i \not=n/2 \}
\cr}
}
(We have simply $USp(v_0)\otimes USp(v_1)$ for $n=2$.)
The conditions on the scalar fields are:
\eqn\hyphomiii{
\eqalign{
X_{01}& = \epsilon_{v_0}  (X_{n-1,0}) ^{tr} \cr
X_{i,i+1} & = (X_{n-i-1,n-i})^{tr} \qquad 1\leq i \leq {n-2\over 2}\cr
X_{n/2, (n+2)/2} & = -\epsilon_{v_{n/2}}  (X_{(n-2)/2, n/2})^{tr}\cr}
}
and similarly for $\bar X$.
The quiver diagram is as in \quivfigii. Note that vertices which
are fixed by the reflection symmetry have group $O(v)$ or $USp(v)$.

\ifig\quivfigiii{ A type I quiver diagram for $X_n$, $n$ even, with
$\chi(g,\Omega)=\xi$.}
{\epsfxsize3.0in\epsfbox{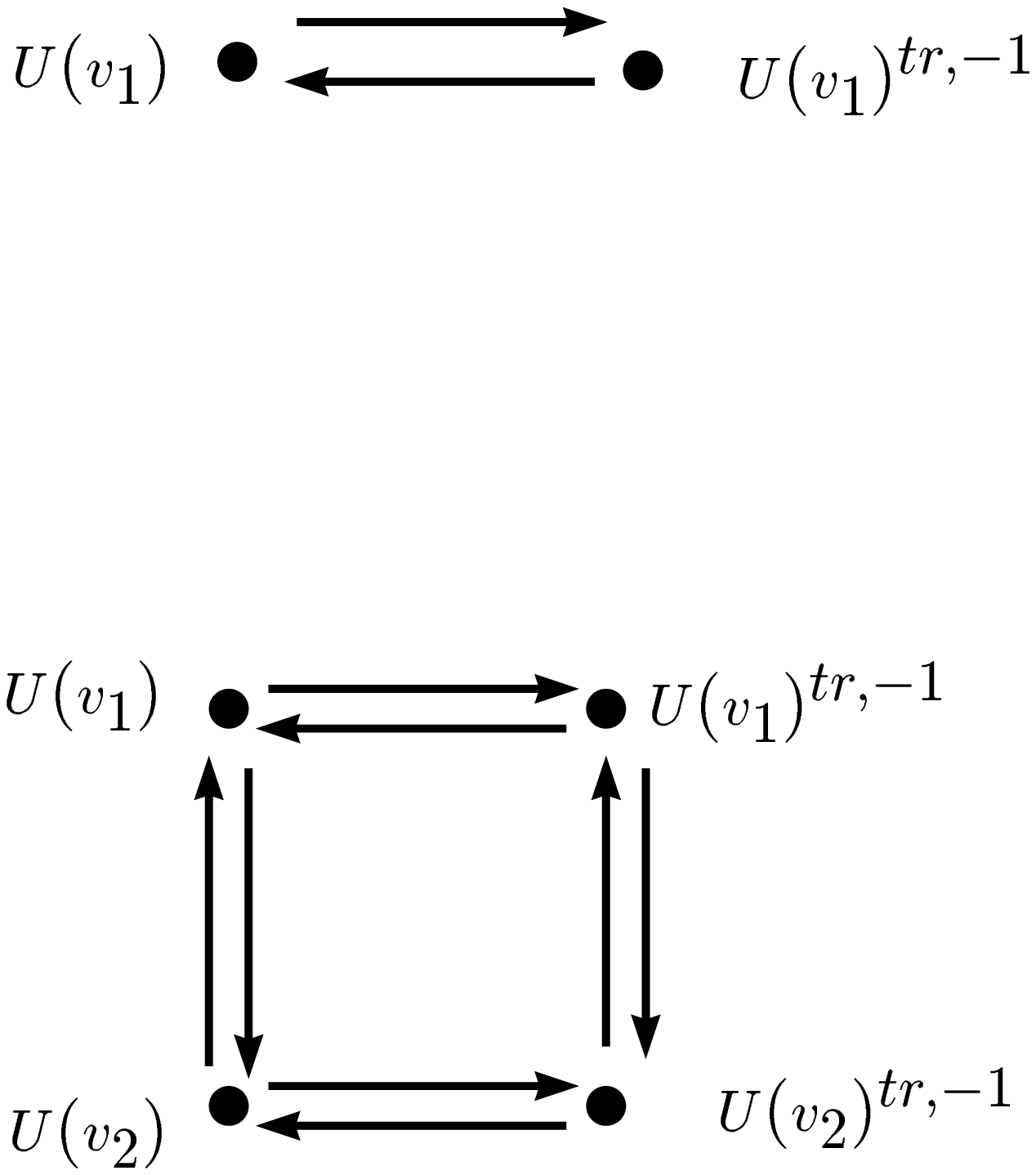}}

\item{I.5.} $\chi(\Omega)=\pm 1, \chi(g,\Omega)=\xi$, (thus $n$ is even).
Here it is more convenient to let the block indices run from
$1$ to $n$ (modulo $n$). We now have:
\eqn\grpii{
\eqalign{
\gamma(\Omega) & =
\pmatrix{
     &     &    & 1_{v_1}    \cr
             &    &   1_{v_2}        &    \cr
             &    \chi(\Omega) 1_{v_2}  &             &        \cr
    \chi(\Omega) 1_{v_1}          &        &         &        \cr}  \qquad
(n=4) \cr
G_5(\vec v) \equiv \bigl[
&
U(v_1) \times U(v_2) \times\cdots U(v_{n/2-1}) \times
U(v_{n/2})\bigr]\cr
\cr
& = \{ (  U_1, \dots, U_{n}) : U_i U_{n-i+1}^{tr} =1 \quad 1\leq i \leq n  \}
\cr}
}
Finally, the conditions on the hypermultiplets are:
\eqn\hyphomiii{
\eqalign{
X_{n,1} ^{tr}& =\chi(\Omega) X_{n1}  \cr
(X_{n-i,n-i+1})^{tr} & = X_{i,i+1} \qquad 1\leq i \leq {n-2\over 2}\cr
 (X_{n/2, (n+2)/2} )^{tr} & =\chi(\Omega)  X_{n/2, (n+2)/2} \cr
\bar X_{1, n } ^{tr}& =\chi(\Omega) \bar X_{1,n}  \cr
(\bar X_{n+1-i,n-i })^{tr} & = \bar X_{i+1,i } \qquad 1\leq i \leq {n-2\over
2}\cr
 (\bar X_{n/2+1, n/2} )^{tr} & =\chi(\Omega)  \bar X_{n/2+1,n/2} \cr}
}
again, reorienting the diagram as in \quivfigiii\  we have
symmetry about the vertical.

\newsec{DD and DN spectrum for  $(p,p+4)$ configurations at the fixed point }

Let us now consider the above theories for
$p=3$ in $p+4=7$ (in type \IIb), $p=4$ in $p+4=8$ (in type \IIa),
and $p=5$ in $p+4=9$ (in type I).
In these cases we can use the language of $d=6$, $\CN=1$ or
$d=4, \CN=2$ SYM to describe the spectrum of the theory on
the world-volume.
Let $w^i=v^{(p+4)}_i$ and $v^i=v^{(p)}_i$. The resulting
low energy field content is again nicely summarized by
quiver diagrams.

\ifig\quivfigvi{ A type II quiver diagram for $(4,8)$ brane configurations
on $X_1=\IR^4$. }
{\epsfxsize3.0in\epsfbox{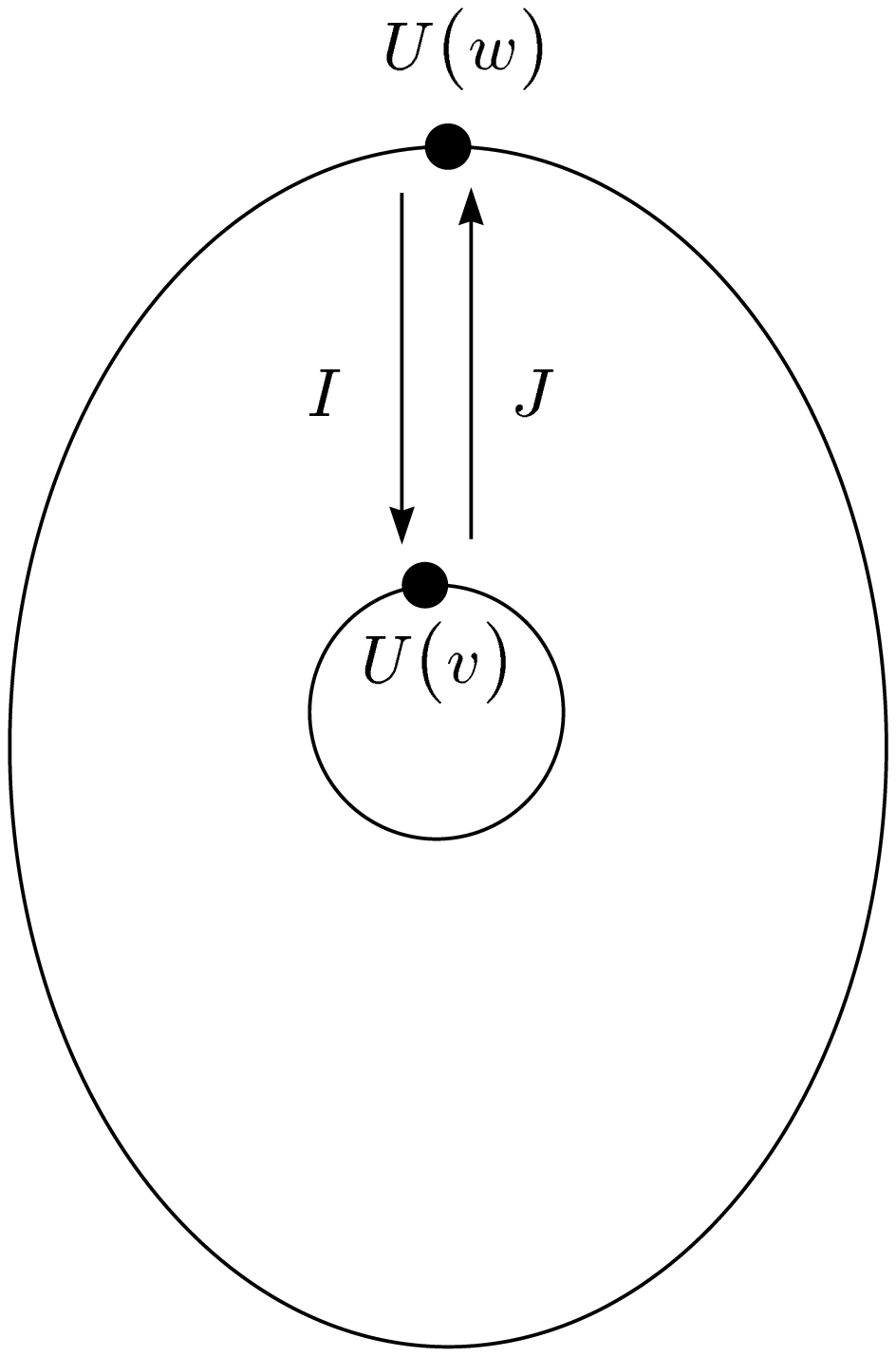}}

\ifig\quivfigiv{ A type II quiver diagram for $(4,8)$ brane configurations
on $X_3$. }
{\epsfxsize3.0in\epsfbox{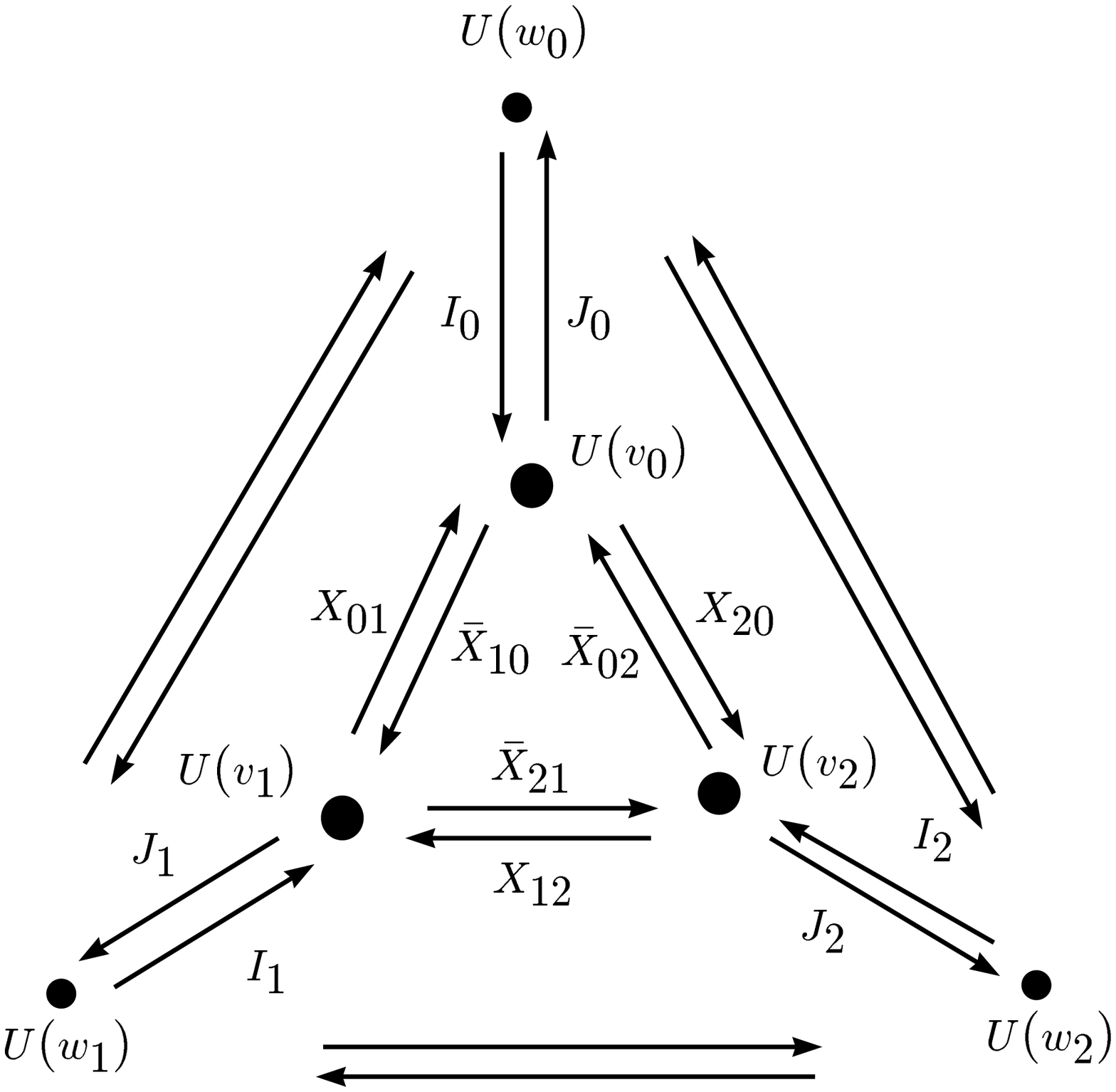}}

There are three sources of fields in the $p$-brane
gauge theory: Restriction of fields from the $(p+4)$-brane,
$p$-brane fields, and $(p+4,p)$-sector fields.
The fields from the $(p+4)$-brane consist of
the restriction
of the vectormultiplets $W_i$. The restriction of
vector fields in the $(p+4)$-brane which are
tangent to $X_n$ gives scalar fields
$Y, \bar Y$ forming hypermultiplets
on the $p$-brane. The $\IZ_n$-projection
requires these to be in:
\eqn\hyphom{
\eqalign{
Y_{i,i+1} & \in Hom(W_{i+1},W_{i})\cr
\bar Y_{i+1,i} &  \in Hom(W_{i},W_{i+1}) \cr}
}
for the $p+4$-brane. These fields comprise the
``outer quiver.'' Note that the  hypermultiplets $Y, \bar Y$
only exist for $p+4<9$.

The fields from the $p$-brane theory
are described as above by an ``inner quiver'' with
vectormultiplets $V_i$ and hypermultiplets
$(X_{i,i+1}, \bar X_{i+1,i})$.
  In addition to this, the inner and outer quivers are joined
by ``spokes'' as in \quivfigvi\quivfigiv.  The spokes correspond
to
the $(p,p+4)$ and $(p+4,p )$ fields $h^{A m}_{~~M}, \tilde h^{A M}_{~~m}$.
The transcription to Kronheimer-Nakajima's notation \kn\nakalg\
is
\eqn\spoketoo{
\eqalign{
J & = \tilde h^1 = (h^2)^\dagger \in \Hom(V,W) \cr
I & = (\tilde h^2)^\dagger= - h^1 \in \Hom(W,V) \cr}
}
Thanks to the reality condition \hyperreality\  we can
  work solely with $\tilde h^1, h^1$ and henceforth we
drop the index $1$.
The $\IZ_n$ projection makes the matrices $I,J$
block diagonal
so that the components are
\eqn\spokes{
\eqalign{
h^{ i m_i}_{~~ i M_i}\leftrightarrow  - I_i    & \in \Hom(W_i, V_i) \cr
\tilde h^{i M_i}_{~~ i m_i}\leftrightarrow   J_i & \in \Hom(V_i, W_i) \cr}
}
 The complete field content is summarized
by the quiver diagrams, e.g., \quivfigvi, \quivfigiv\ give the
diagrams for $X_1, X_3$ respectively.

\subsec{Type \I}

\ifig\quivfigvii{ A type \I\ quiver diagram for $(5,9)$ brane configurations
in $X_1$}
{\epsfxsize3.0in\epsfbox{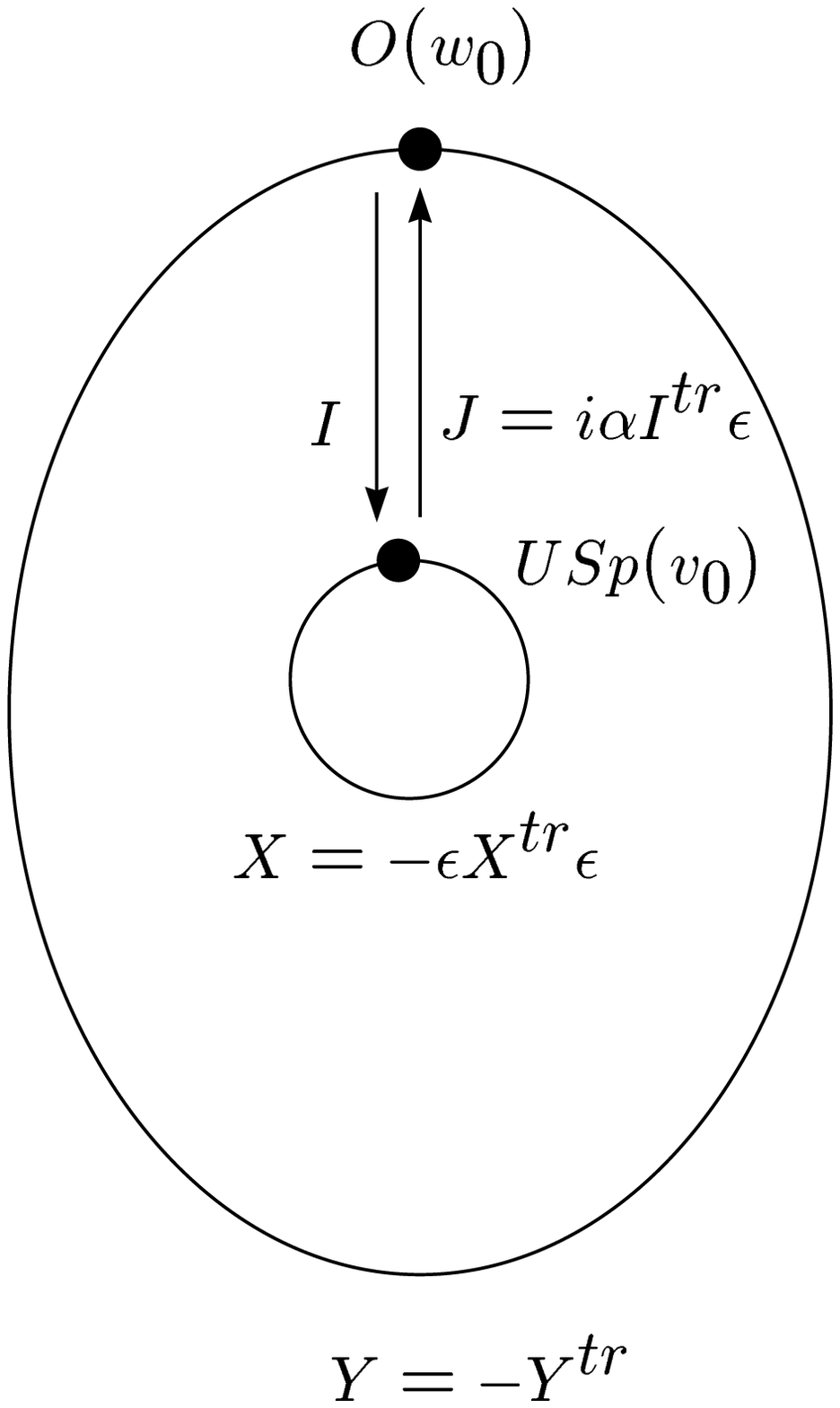}}

\ifig\quivfigv{ A type \I\ quiver diagram for $(5,9)$ brane configurations
in $X_3$}
{\epsfxsize3.0in\epsfbox{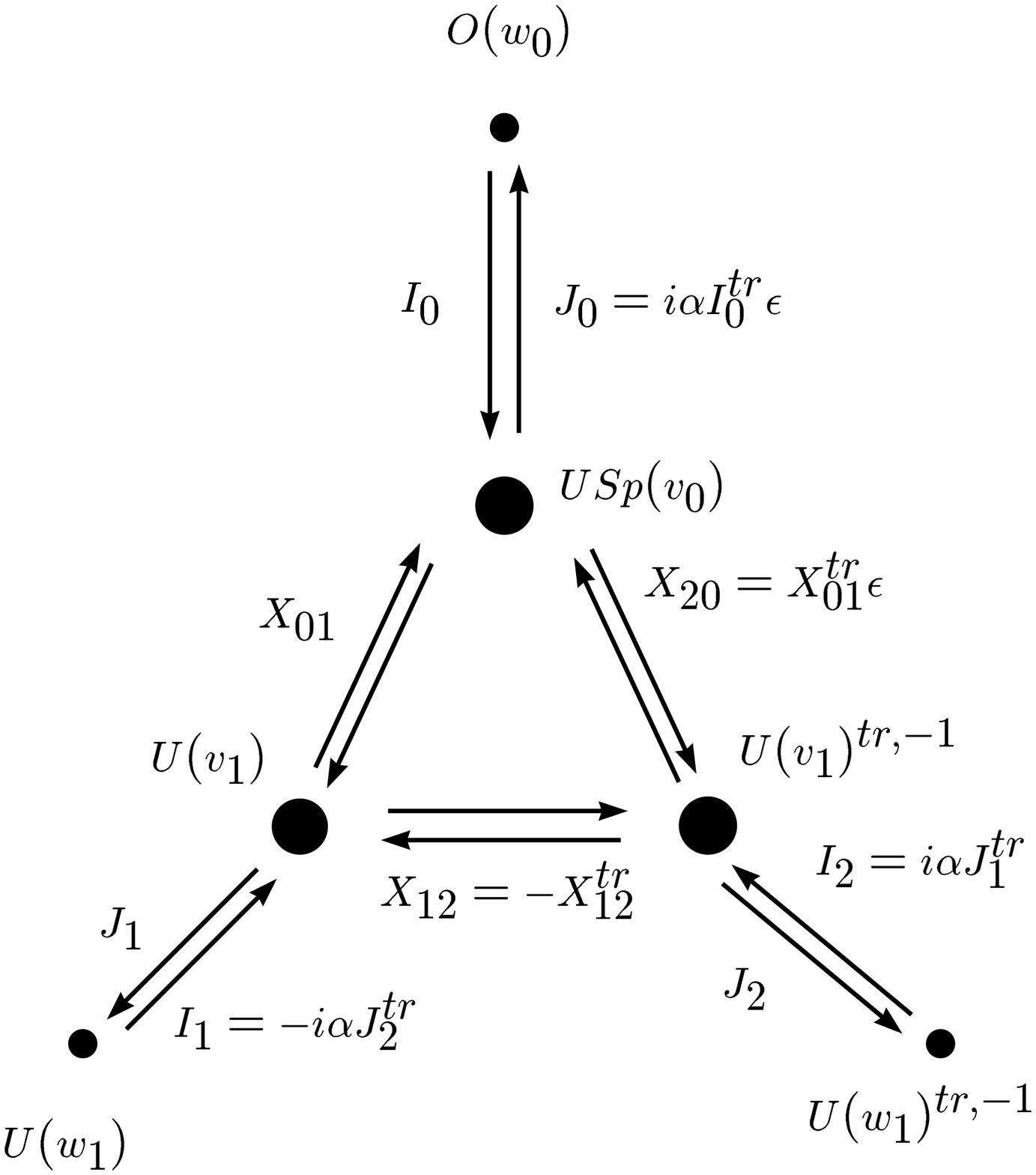}}

\ifig\quivfigviii{ A type \I\ quiver diagram for $(5,9)$ brane configurations
in $X_4$, for $\chi(g,\Omega)=+1$.
 Arrows between the outer dots, associated to the
$(Y, \bar Y)$ hypermultiplets, have been omitted. }
{\epsfxsize3.0in\epsfbox{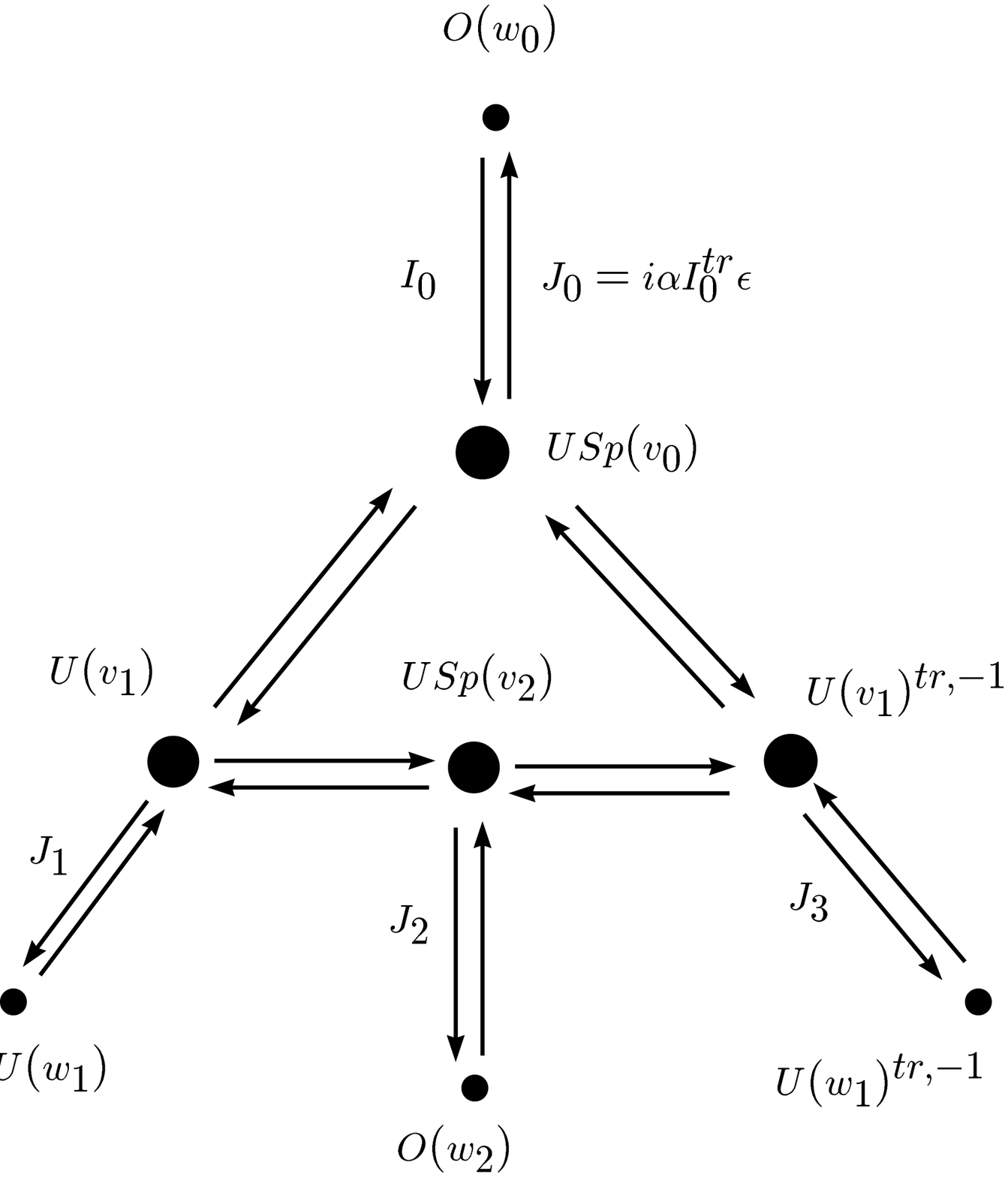}}

The conditions for  \om\ on the inner hypermultiplets $(X, \bar X)$
have been described in detail in the previous section.
The conditions on the outer hypermultiplets $(Y, \bar Y)$
have an extra sign change relative to the condition for $X$:
\eqn\signsut{
Y =- \gamma_9(\Omega) Y^{tr} \gamma_9(\Omega)^{-1}
}
since they are restrictions of gauge fields.
The   conditions    \om\ are:
\eqn\jaytr{
J^{tr} = - i   \gamma_5(\Omega) I \gamma_9(\Omega)^{-1}
}

There are several type \I\ quivers depending on the various
unbroken groups $G_i(\vec w)$ and $G_i(\vec v)$ we associate
to the inner and outer quivers.
We will consider just two cases

I. $\chi_9(\Omega)=+1, \chi_5(\Omega)=-1, \chi_9(g,\Omega)=
\chi_5(g,\Omega)=1$.

\eqn\hrealty{
\eqalign{
J_0^{tr} & = - i   \epsilon_{v_0} I_0 \cr
J_k^{tr} & = - i   I_{n-k} \qquad 0< k<n/2 \cr
J_k^{tr} & = + i   I_{n-k} \qquad n/2< k<n  \cr
J_{n/2}^{tr} & = - i   \epsilon_{v_{n/2}} I_{n/2} \qquad n\  {\rm even} \cr}
}
The field content is summarized by the  quivers shown in
\quivfigvii, \quivfigv, \quivfigviii.

\ifig\quivfigix{ A type I quiver diagram for $(9,5)$-brane configurations
on the Eguchi-Hanson space $X_2$ with $\chi_9(g,\Omega)=\xi=-1$. }
{\epsfxsize3.0in\epsfbox{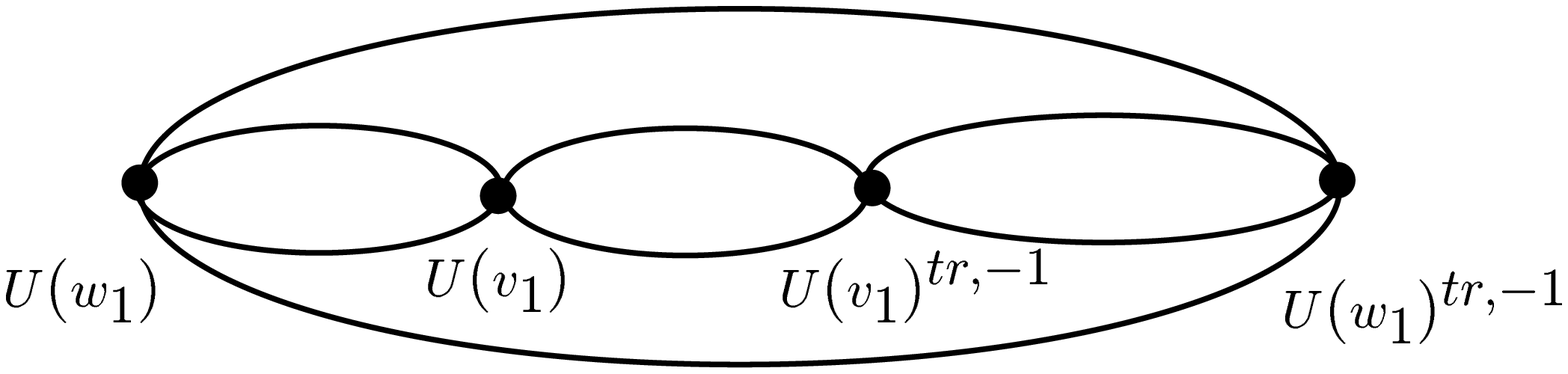}}

\ifig\quivfigx{ A type I quiver diagram for $(9,5)$-brane configurations
on   $X_4$ with $\chi_9(g,\Omega)=\xi=i $.}
{\epsfxsize3.0in\epsfbox{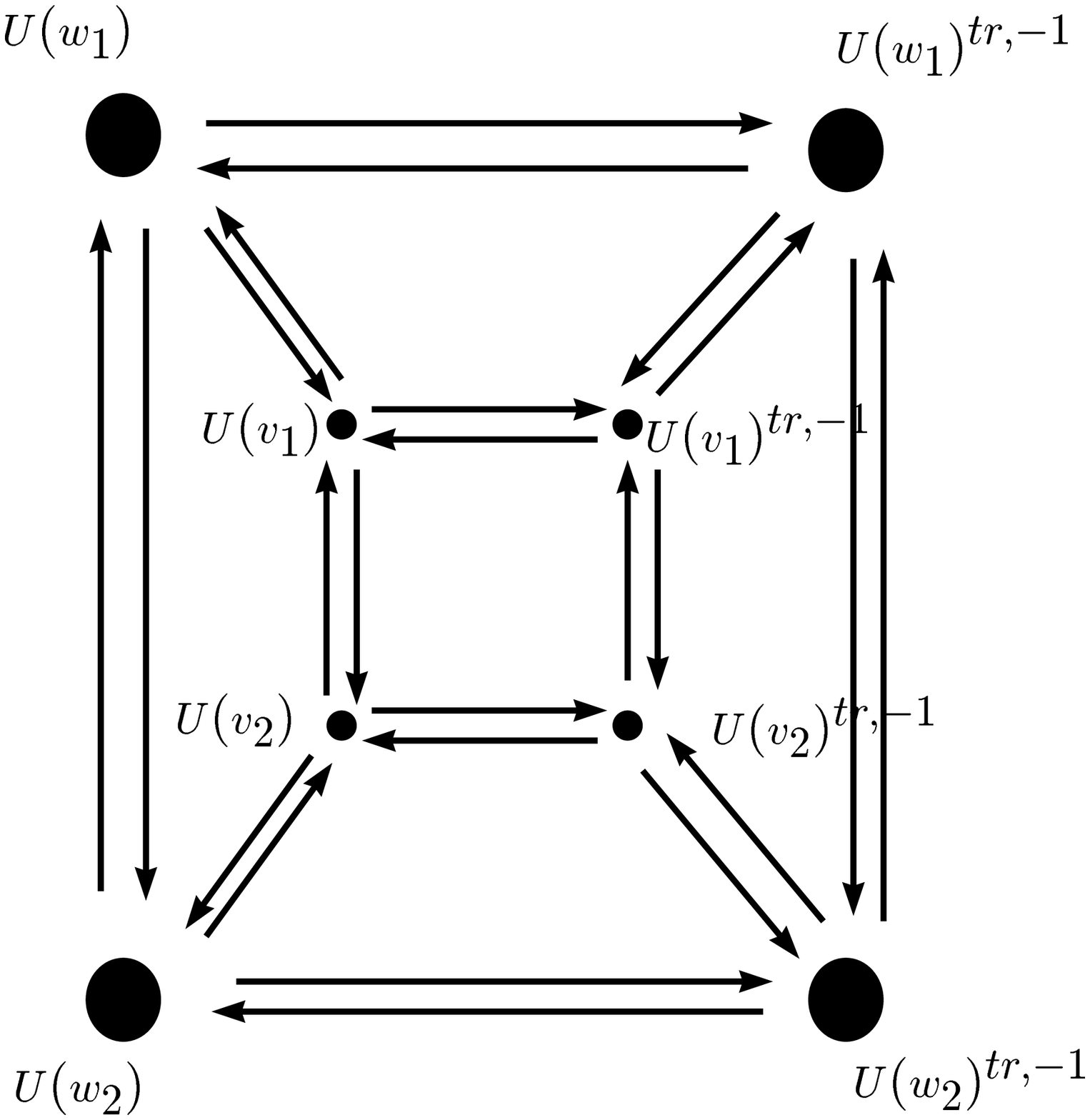}}

II. $\chi_9(\Omega)=+1, \chi_5(\Omega)=-1, \chi_9(g,\Omega)=
\chi_5(g,\Omega)=\xi$.

This case can only occur when $n$ is even.
Letting indices run from $1 $ to $n$ the conditions on the hypermultiplets
joining the inner and outer quiver are
\eqn\spokethree{
\eqalign{
J_j^{tr} & = i   I_{n+1-j} \qquad j\leq n/2 \cr
J_j^{tr} & = - i   I_{n+1-j} \qquad j >  n/2 \cr}
}
The spectrum summarized by \quivfigix\ is that worked out
by Gimon and Polchinski \gimon.

\newsec{World-volume action for $p$-branes transverse to the
fixed point}

Having described the world-volume spectrum (at the fixed point)
in sections 2,4,5 we now proceed to describe the Lagrangian
governing the low energy dynamics. There is no simple
unified formulae for D-brane actions yet, but several terms
are now well-known:
\eqn\sevterms{
I = I_{BI} + I_{HM} + I_{CS} +I_{\rm susy} + \cdots~.
}
$I_{BI}$ is the Born-Infeld action
$\int_{\CB_p \times \IR} \Tr \sqrt{\det (G+ \CF)} $
where $\CF= F-B$. Expanding the squareroot gives the
Yang-Mills action at leading nontrivial order. $I_{HM}$
gives the kinetic energies of the hypermultiplets.
$I_{CS}$ is a Chern-Simons coupling found in
\douglas, $I_{susy}$ contains the supersymmetric
completions of the lowest order terms and $\cdots$ hides
our ignorance about higher order terms in the low-energy
expansion.  In this section we describe in some detail
$I_{CS}$ and $I_{\rm susy}$ for the $5,4,3$-brane
in the type $\I,\IIa,\IIb$ theories.

The Chern-Simons couplings are described in
general as follows: Let $C$ denote the sum of
$p$-form fields (in ten dimensions). Then, for
a flat  $D$-brane in $\IR^{10}$ we have:
\eqn\csi{
I_{CS} = \int_{\CB_p \times \IR}  C\wedge  \Tr e^{\CF}
}

Now let us consider the modifications in
the presence of an orbifold. Of course, we
retain \csi\ where $C$ comes from the
untwisted sector.
The definition of D-branes on an orbifold correlates the action
of a point group element   on the world-sheet and Chan-Paton factors,
so that the closed string sector twisted by $g$ will couple to an open
string boundary with Chan-Paton factors twisted by $\gamma(g) $.
Thus we expect extra Chern-Simons couplings:
\eqn\csit{
I_{CS} = \int_{\CB_p \times \IR}  \sum_{k=1}^{n-1}
{}~~
{}^{p+1}C_k \wedge \Tr \gamma(g^k) e^{\CF} .
}
The RR fields ${}^{p+1}C_k$ are a bispinor
field in the $k$-twisted sector, restricted to the $(p+1)$-dimensional
world-volume.
The existence of these couplings is checked
by a vertex operator calculation in appendix A.
Additional couplings to hypermultiplet scalars are obtained by
the replacement
$\CF \rightarrow \CF + d X^i b_i$ explained
in \douglas.

As described in the appendix, \csit\ is exact only when the D-branes
are coincident with the orbifold fixed point.  At non-zero distance
$|X|$, we expect this coupling to be suppressed as
$\exp - |X|^2/\alpha'$.
We will neglect this here, obtaining results valid for $|X|^2<<\alpha'$.

\subsec{\Hk moment map}

Let us now consider the terms in the
supersymmetric completion involving only
bosonic fields.
The completion of $I_{BI}+ I_{HM}$ in 6d
involves coupling the
triplet of $D$-terms for
$d=6, \CN=1$ SYM to the hypermultiplets
through the \hk moment map.

Quite generally, in a linear $d=6,\CN=1$ or $d=4, \CN=2$ theory
the hypermultiplets
are described
by starting with a complex hermitian vector space $V$ with
a unitary action of the gauge group $G$.
The hypermultiplets take values in
the  vector space $V\oplus V^*$. This space is
a quaternionic vector space. Indeed,
if we choose
coordinates:
$  \{ z^\alpha \}_{\alpha=1, \ell} $ for $V$ and
 dual coordinates $w_\alpha$ for $V^*$ then
we define quaternionic coordinates as in subsection 2.2,
\eqn\quatthree{
\BX^\alpha  = \pmatrix{ z^\alpha & \bar w^\alpha\cr
-w_\alpha & \bar z_\alpha\cr}
}
so that the complex structures $\cI,\cJ,\cK$ correspond to
right multiplication by $i\sigma_3, i\sigma_2, i \sigma_1$,
respectively.
Moreover,
$G$ acts via
\eqn\iv{
\delta_\Lambda (z^\alpha ; w_\alpha)=
\{(T_\Lambda)^\alpha_{~~ \beta} z^\beta ;
- w_\alpha (T_\Lambda)^\alpha_{~~ \beta} \}
}
where
  $ 1\le \Lambda\le\dim G$ is an index labelling a
basis for $\lieg$, and
$ [ (T_\Lambda)^\alpha_{~ \beta} ]^* = - (T_\Lambda)^\beta_{~ \alpha} $.
This action may be written as:
\eqn\hkacti{
\delta_\Lambda \BX^\alpha  = (\tau_\Lambda)^\alpha_{~~ \beta} \BX^\beta
}
where we replace $T$ by $\Re T\cdot\cone + \Im T\cdot\cI$:
\eqn\hkactii{
(\tau_\Lambda)^\alpha_{~~\beta} =
 \pmatrix{(T_\Lambda)^\alpha_{~~ \beta}&0 \cr
0& \bigl[ (T_\Lambda)^\alpha_{~ \beta} \bigr]^* \cr} .
}
In general hypermultiplets take values in
a \hk manifold. For the vector space $V\oplus V^*$
we have Kahler and holomorphic symplectic forms
\eqn\kps{
\eqalign{
\omega^R & = {i\over 2}\sum [dz^\alpha d\zb_{\alpha} +
dw_\alpha d\wb^{\alpha}] \cr
\omega^C & = \sum dz^\alpha\wedge dw_\alpha\cr}
}
these forms comprise a triplet $\vec \omega$.
The $G$-action is symplectic with respect to
each of these forms and hence we obtain a
triplet of Noether charges
defining the \hk\ moment map:
\eqn\nkmm{
\vec\mu_\Lambda = \half\tr \vec\sigma \BX^\dagger_\alpha
	(\tau_\Lambda)^\alpha_{~ \beta} \BX^\beta \qquad .
}
Explicitly:
\eqn\qautmm{
\eqalign{
  \mu_\Lambda^R& = \half\bigl[ \zb_\alpha (T_\Lambda)^\alpha_{~\beta} z^\beta
- w_\alpha (T_\Lambda)^\alpha_{~~ \beta}\wb^\beta\bigr] \in \sqrt{-1} \IR \cr
\mu_\Lambda^C & = w_\alpha (T_\Lambda)^\alpha_{~\beta} z^\beta \cr}
}

The completion of $I_{BI}+ I_{HM}$ in 6d  is:
\eqn\suscp{
\int_{\CB_5 \times \IR}  \sum_\Lambda
\bigl[
(\vec D_\Lambda)^2+ \vec D_\Lambda \cdot \vec \mu_\Lambda
\bigr]
}
Here the sum runs over a basis $\Lambda$ for the
Lie algebra of the worldbrane gauge group
and $\vec D_\Lambda$ is a triplet of auxiliary D-fields in
the vectormultiplet.
For $p=3,4$ we simply reduce \suscp. In particular,
for $p=3$
we obtain the D and F - auxiliary fields:
$D_\Lambda= D_\Lambda^r, F_\Lambda = D_\Lambda^c$.

The completion of the Chern-Simons couplings
is trickier, in part because they are {\it not} obtained by
naive dimensional reduction but rather by re-applying \csi.
This is ultimately because
the surviving supersymmetry $\tilde\eps=\Gamma_D\eps$ is different for
each $p$.
We turn to this next.

\subsec{$IIb$}

Here we take $p=3$.
We must reduce the 6d spectrum of  section 2.4
 to $3+1$ dimensions.  Let us take coordinates
$0,1,2,3$ along the 3-brane $\CB_3\times \IR$
and coordinates $6,7,8,9$ to describe the
ALE space.
The twisted sector matter
 multiplets decompose into  a sum of  a
linear hypermultiplet
$( {}^{4}C^{(2)}, \vec {\tilde \phi}_k)$
and a vector-multiplet of
$d=4, \CN=2$.
\foot{Linear hypermultiplets are
described in the next section.}
The vectormultiplet may be taken to
be
 $(({}^{4}C^{(1)}_k)_{5 \mu} , {}^{4}C^{(0)}_k, b_k^{(0)} ) $
where the vector fields $({}^{4}C^{(1)}_k)_{5 \mu} $  and
$({}^{4}C^{(1)}_k)_{4 \mu} $ are related by $d=4$ duality.

The twisted sector couplings become
\eqn\csiib{
\eqalign{
\int_{\CB_3 \times \IR}
&
{}^{4}C^{(0)}_k \Tr \gamma(g^k) \CF \wedge \CF
+ b^{(0)}_k \Tr \gamma(g^k) \CF \wedge *\CF \cr
& +
[(X^4 + i X^5) {}^{4}H_k^{(2)+} +  (X^4 - i X^5) {}^{4}H_k^{(2)-}]
	\Tr \gamma(g^k) \CF
+
{}^{4}C^{(2)}_k \Tr \gamma(g^k)   \CF \cr}
}
Here ${}^{4}H^{(2)\pm}_k$ are the
field strengths of $({}^{4}C^{(1)}_k)_{5 \mu}$;
we have integrated by parts and discarded a term
higher order in derivatives.

We now discuss the supersymmetric completion of the
terms \csiib. The first two terms contribute to standard
couplings of vectormultiplet gauge fields to vectormultiplet
scalars.  The last term is somewhat more unusual,
and gives a coupling between the hypermultiplet scalar
${}^4C^{(2)}_k$ and vectormultiplets. Its supersymmetric
completion involves the NS scalars $\vec {\tilde \phi}_k$.
As described in more detail in the next section
this completion is the $d=6, \CN=1$
Fayet-Iliopoulos term:
\eqn\csiibi{
\sum_{k=0}^{n-1}
\int_{\CB_3 \times \IR}
\vec {\tilde \phi}_k  \Tr  \gamma(g^k) \vec D
}
Note, in particular, that the sum  on $k$ includes the
untwisted sector.
The existence of these couplings -- which play a crucial role in what
follows -- may be deduced from world-brane supersymmetry.
As we show in detail in the following section, they may also
be predicted from an analysis of anomalies. Finally, it is
relatively straightforward to verify their existence by
an explicit  vertex operator calculation. This is done
in appendix A.

\subsec{$IIa$}

Here we take $p=4$.
Now we must reduce the 6d spectrum of section 2 to 5d.
The untwisted matter multiplet and $(n-1)$ twisted
matter multiplets again reduce to a vm + hm,
now $({}^{5}C^{(1)},b_k^{(0)} ) $ and $( {}^{5}C^{(3)}, \vec {\tilde \phi}_k)$.

{}From the untwisted sector we have
\eqn\iiacsi{
\int_{\CB_4 \times \IR}  {}^{5}C^{(1)} \Tr \CF\wedge \CF +
{}^{5}C^{(3)} \Tr \CF + v {}^{5}C^{(5)}
}
while the twisted sector contributes
\eqn\iiacsi{
\sum_{k=1}^{n-1}
\int_{\CB_4 \times \IR}  {}^{5}C^{(1)}_k \Tr \gamma(g^k)  \CF^2
+ b^{(0)}_k \Tr \gamma(g^k) \CF \wedge *\CF
+ {}^{5}C^{(3)}_k \Tr \gamma(g^k) \CF
}
Upon dimensional reduction to $3+1$ we recognize the
standard couplings of vector multiplet scalars to
gauge fields governed by a prepotential.
The ${}^{5}C^{(3)}_k \CF$ coupling reduces to the ${}^{4}C^{(2)}_k \CF$
coupling described above, and again supersymmetry will require the
Fayet-Iliopoulos coupling \csiibi.

\subsec{Type \I }

Each twisted sector gives rise to a hypermultiplet
$({}^{6}C^{(4)}_k,\vec{\tilde\phi_k})$.
The $5$-brane twisted sector Chern-Simons coupling is
\eqn\icsi{
\sum_{k=1}^{n-1}
\int_{\CB_5 \times \IR} {}^{6}C^{(4)}_k \Tr \gamma(g^k) \CF     .
}
and again its partner Fayet-Iliopoulos couplings \csiibi\ are present.

The couplings to the $9$-brane are given by a simple modification
to the world-sheet calculations of the appendix.
They are similar to \csiib\ but localized at the fixed points $x_i$,
\eqn\ggstn{\int\ d^{10}x\ \delta^{(4)}(x-x_i)\wedge {}^{6}C^{(4)}_k \wedge
\tr \gamma(g^k)\ F,}
Here ${}^{6}C^{(4)}_k$ is the $4$-form dual to the type I scalar RR
field in the $k^{th}$ twisted hypermultiplet.  There is no such
coupling in the untwisted sector.

The NS-NS partner of this term is the same as \csiibi,
except that there is no contribution from the untwisted sector
$k=0$.

\subsec{Hypermultiplet potential energy}

Finally, let us work out the potential terms for the
scalars in the hypermultiplets.  For definiteness
we work in the \IIb\  theory.
We would like to integrate out the $\vec D_\Lambda$
in \suscp. Therefore, we must diagonalize the
couplings \csiibi.

Let $F_{j}$ be the field strength of the $U(1)$ factor contained in $U(v_j)$,
and let $\tilde C_i = {}^{4}C^{(2)}_i$ be the RR potential in the $g^i$ twisted
sector; then
we may rewrite the Green-Schwarz coupling  from \csiib\  as
\eqn\ggsts{
\int_{\CB_3 \times \IR}   \sum_{i,j=0}^{n-1} \xi^{i\cdot j}\tilde C_i\wedge
F_j.
}

The couplings can be diagonalized by doing a discrete Fourier transform
on either $F_j$, to produce $\tilde F_i$ or on $\tilde C_i$ to produce $C_j$.
On the one hand, $\tilde F_i$ is the $U(1)$ gauge factor in the ``$i$'th
twisted open string
sector,'' connecting a D-brane with its image under $\gamma(g) ^i$.
On the other hand, $C_j$ couples to the $U(1)$ in a single factor $U(v_j)$.

It will turn out (in section 8) that the $C_j$ have a simpler
interpretation (they are associated with individual two-cycles),
so let us use this basis from now on, and rewrite \ggsts\ as
\eqn\ggsttwo{\int_{\CB_3 \times \IR}  \ \sum_j C_j\wedge F_{j}.}
$(0,1)$ supersymmetry on the world-volume requires the partners
\eqn\ggstsum{
\int_{\CB_3 \times \IR}  \  d^4 x\ \sum_j \vec\phi_{j} \cdot   \vec D_j
}
Here $\vec  D_j$ is the contribution
of D-brane matter to the \hk\ map for
the  $U(1)$ in $U(v_j)$.
%
%

We are finally able to complete the square to get the
hypermultiplet potential energy. Define
$ \vec \phi_\Lambda$ to be zero for noncentral
generators of $G(\vec v)$ and
$ \vec \phi_\Lambda = \vec \phi_j$ if $T_\Lambda$
is the $U(1)$ generator of $U(v_j)$.
Integrating out
$\vec D_\Lambda$ we have simply
\eqn\poteng{
\sum_\Lambda (\vec \mu_\Lambda - \vec \phi_\Lambda)^2
}
In what follows we will denote
the vev's of these scalars by
\eqn\zetas{
\langle \vec \phi_\Lambda \rangle \equiv \vec \zeta_\Lambda
}

\newsec{Anomalous $U(1)$'s and Fayet-Iliopoulos terms}

\subsec{$U(1)$ anomaly cancellation}

Another way of understanding the presence of the
crucial couplings \csiibi\ is through anomaly cancellation.
In  $d=6,\CN=1$, theories with charged $U(1)$ fields are always anomalous.
This is because supersymmetry correlates the type of supermultiplet,
vector or hypermultiplet, with the chirality of the fermions they
contain.  Since only hypermultiplets can have $U(1)$ charge, all
contributions to the $F^4$ anomaly will have the
same sign.

Several people (we learned it from John Schwarz)
have noticed that the theory of Gimon and Polchinski is
an example, and that a $d=6$ version of the mechanism proposed for
$d=4$ by Dine, Seiberg and Witten \dsw\ will resolve the problem.
The idea is that the couplings required for $d=10$ Green-Schwarz anomaly
cancellation, when evaluated with non-zero background gauge field in the
internal space, will lead to couplings of the form
\eqn\gsone{\int\ d^Dx\ C^{(D-2)} \wedge F,}
where $F$ is the $U(1)$ field strength and $C^{(D-2)}$ is a $D-2$-form
gauge field.  Cancellation of the $F^{(D-2)/2}$ axial anomaly
requires a gauge transformation law
$C^{(D-2)} \rightarrow
C^{(D-2)} + \epsilon F^{(D-2)/2}$.
The added term in the world-volume action changes the
duality transformation to a scalar $c^{(0)}$
to
\eqn\dltytm{
*_{D} d c^{(0)} = H^{D-1} + *_D A
}
and hence  $c^{(0)}$ has  a non-trivial gauge
transformation:
\eqn\gaugetr{\delta A_\mu = \p_\mu \epsilon \qquad\qquad
\delta c^{(0)} = \epsilon}
and couplings
\eqn\gaugecou{\int\ d^Dx\ (A_\mu - \p_\mu c^{(0)})^2.}
The resulting theory is gauge-invariant and describes a massive vector boson.

The story becomes even more interesting when supersymmetry is taken
into account.  In $\CN=1$, $d=4$ supersymmetric Yang-Mills theory, the
scalar $c^{(0)}$ is one real component of a chiral superfield
$C$ with lowest component $c^{(0)}+i\phi$.
Let $V$ be the vector superfield containing $A_\mu$; then
a superfield coupling containing \gaugecou\ is
\eqn\supgaugecou{\int\ d^4x\ d^4\theta\ \Im\ {1\over 4}(C-\bar C-V)^2.}
This is gauge invariant if $\delta V=\Lambda-\bar\Lambda$
and $\delta C=\Lambda$.
It also produces a coupling to $\phi$,
\eqn\dterm{-\int d^Dx\ \phi\ \Im\ \int d^4\theta\ V.}
Thus $\phi$, the partner to $c^{(0)}$, controls the Fayet-Iliopoulos term.

More general K\"ahler potentials $K(C-\bar C-V)$ are allowed but are better
discussed in the language of subsection 7.3.

\subsec{Generalization to $\CN=2$, $d=4$}

$\CN=1$, $d=6$ $U(1)$ gauge theories and their dimensional reductions
can have three Fayet-Iliopoulos terms
$\vec\zeta=(\Re\zeta^C,\Im\zeta^C,\zeta^R)=\vec\sigma_{AB}\zeta^{AB}$,
forming a triplet of $SU(2)_R$.
Let us explain this in the (perhaps more familiar)
$\CN=2$, $d=4$ case, using $\CN=1$ superfield notation and assembling
the vector multiplets as $(V,A)$.
The superpotential is almost
uniquely determined by the gauge group $G$ and matter representation $R$.
(Mass terms are allowed but are not relevant for us.)
The conditions for a supersymmetric vacuum
for $U(1)$ gauge theory with hypermultiplets $(M_i,\bar M_i)$
of charge $(q_i,-q_i)$ are\footnote*{
The additional conditions $\p W/\p M_i=0$ are not present in $d=6$
(where $A$ becomes the $5$ and $6$ gauge field components), and
in any case are not relevant for describing moduli
spaces of instantons on the ALE space.}
\eqn\ffourtwo{\eqalign{
0 &= D = -\zeta^R + \sum_i q_i (M^*_i  M_i - \bar M^*_i \bar M_i)\cr
0 &= F = {\p W\over\p A}\cr
W &= -\zeta^C A + \sum_i q_i \bar M_i\ M_i A.}}
The term $\zeta^C A$ is an $SU(2)_R$ covariant
generalization of the $\CN=1$ FI term.

These equations can also be written in an $SU(2)_R$ covariant form,
using `quaternionic' notation (as in \quatone,\quatthree)
\eqn\quattwo{
\BM^{A'A}  = \pmatrix{ M & \bar M^*\cr
-\bar M & M^*\cr}
}
satisfying $(\BM^*)_{A'A} = \eps_{A'B'}\eps_{AB}\BM^{B'B}$.
Then
\eqn\defmu{
\vec\zeta = \vec\mu_j = \half\vec\sigma_{A}^{~B}~~
 \sum_i q_i\ \BM_i^{A'A} (\BM_i^*)_{A'B}.
}

As with $\CN=1$, such Fayet-Iliopoulos terms will be $\CN=2$ supersymmetry
partners of the anomaly cancelling coupling \gaugecou.
Now $c^{(0)}$ will be one component
of a hypermultiplet also containing $\vec\zeta$,
which we write in terms of $\CN=1$ chiral superfields as $(C,\Phi)$.
The $\CN=2$ extension of \supgaugecou\ is simply
\eqn\suptwogaugecou{
\int\ d^4x\ d^4\theta\ \Im\ {1\over 4}(C-\bar C-V)^2 + \Phi\bar\Phi
+ \int\ d^4x\ d^2\theta\ \Phi\ A + {\cc}.
}

\subsec{Linear hypermultiplets}

In a general string compactification,
the kinetic terms for the moduli need not take the form \suptwogaugecou.
For example, the gravitational moduli for type \II\ on K3 live on
a homogeneous space.

The mechanism just described works for more general kinetic terms.
It is best
described in terms of a ``linear hypermultiplet'' whose components
are the R-R field strength $H^{(D-1)}$ and the three NS-NS scalars.
Its component fields correspond directly to the world-sheet vertex operators,
and form a $1+3$ of $SU(2)_R$, as we saw for the moduli in section 2.3.
A $d=6$, $\CN=1$ superfield version of the multiplet is given in \hst.

These NS-NS scalars are always equal to the FI terms $\vec\zeta$.
On the other hand, the relation to the standard hypermultiplet is through
dualizing $H^{(D-1)}$, which for a general kinetic term is a non-linear
transformation.  In this case the $\phi$ and $\Phi|_{\theta=0}$
of the previous section will not be equal to $\vec\zeta$.

In terms of $d=4$, $\CN=1$ superfields the multiplet becomes a
``linear chiral multiplet'' $G$ containing a $3$-form field strength and
a scalar, and a chiral multiplet $\eta$ containing the other scalars.
The Lagrangian dual to \suptwogaugecou\ is \superspace
\eqn\gssup{
\int d^4x\ d^4\theta\ G~V + \int d^4x\ d^2\theta\ \eta~A + {\rm c.~c.}
}
combined with the kinetic term
\eqn\getafree{
\int d^4x\ d^4\theta\  \biggl( -\half G^2 + \eta\bar\eta \biggr).
}
A more general kinetic term
\eqn\getagen{
\int d^4x\ d^4\theta f(G,\eta,\bar\eta)
}
will be supersymmetric if $f$ satisfies
\eqn\getafcon{
\left({\p^2\over\p G^2} + {\p^2\over\p \eta\p\bar\eta}\right)f=0.
}
Dualizing $G$,
a transformation described explicitly in \refs{\hitchin,\superspace},
produces the hypermultiplet form of the theory.
All this generalizes to $n$ linear multiplets and provides a general
construction of $4n$-dimensional \hk metrics with $U(1)^n$ symmetry.

\subsec{Application to the D-brane theories}

It should now be clear that anomaly cancellation for the
\IIb\ theory requires the couplings \csiib\csiibi.
We may also develop the formal realization of this mechanism for
a type \IIb\ compactification on $\BC^2/\BZ_n$ with $9$-branes and
$5$-branes at the fixed point.
Although this theory is anomalous, we will be able to derive
the type \I\ result from this by applying the $\Omega$ projection,
and the sensible type \II\ results by dimensional reduction.

A $5$-brane will have couplings to the Ramond-Ramond potentials $C$
of the generalized Green-Schwarz form described in \douglas,
including the term
\eqn\ggs{\int\ d^6x\ C^{(4)} \wedge \tr F.}
This will serve to cancel the anomalous $U(1)$ contained in $U(v_0)$.
However, our orbifold theories contain numerous $U(1)$'s,
$n$ for type \II, each of which requires its own $c^{(0)}$ for
gauge invariance.
These come from Ramond-Ramond $4$-form potentials in the twisted sectors
as above.

The $9$-brane $U(1)$ gauge theories are also anomalous,
with part of the anomaly from $9$-brane matter $Y$,
and part from $5$-brane matter $H$.
The $5$-brane couplings are derived from the results of section 6.4
by imposing the $\Omega$ projection.  Applying the projection to the
$5$-brane matter in fact eliminates couplings to the type \IIb\ closed
string fields not present in type \I; for example the $U(1)$ in $U(v_0)$
is removed, eliminating \ggs, which is consistent with the removal of
$C^{(4)}$.

A pair of twist sectors related by $\Omega$ as in
section 4.2, e.g. $i$ and $n-i$, will be related to one $9$-brane $U(1)$
and one $5$-brane $U(1)$.  For the anomaly cancellation to work for
both $U(1)$'s \csiib\ and \ggstn\ must couple to different linear combinations
of $C_i$ and $C_{n-i}$.  This is true (and essentially follows from
\relchis).

\newsec{D-flatness equations and the moduli of D-brane ground states}

In this section we write the equations determining the
collective coordinates of D-branes transverse to the
ALE space realized as a blown-up orbifold.

According to \poteng\ we must solve the
equations
\eqn\dflat{
\vec \mu_\Lambda = \vec \zeta_\Lambda\qquad \mod G
}
 Recall that
$\vec \zeta_\Lambda$ is only nonzero for central
generators of the gauge group.

The equations \dflat\ define a special case of
a general construction -- the \hk quotient.
 Hyperkahler quotients have been discussed extensively
(see for example \refs{\hitchini,\hitchin}).
Quite generally,
if $X$ has a \hk metric, then the tangent space
admits an action by the quaternions $\IH$, and $X$ has
3 covariantly constant symplectic forms which may
be assempled into a vector $\vec \omega$ in the
Lie algebra of the unit quaternions $sp(2)$. If a Lie
group $G$ with Lie algebra $\lieg$
acts on $X$ preserving the \hk structure
then for all $\xi\in \lieg$ we have
$d\vec \mu(\xi) + \iota_\xi \vec \omega=0$.
The function $\vec \mu(\xi) $ is the Noether charge,
and defines a map $\vec \mu: X \rightarrow \lieg^*\otimes sp(2)$.
If $\vec \zeta\in Center(\lieg)^*\otimes sp(2)$ then the level
set $\vec \mu^{-1}(\vec \zeta)$ is $G$-invariant, an we may take
the quotient: $(X/G)_{\vec \zeta } \equiv \vec \mu^{-1}(\vec \zeta)/G$.
If we choose a complex structure on $X$ then the the \hk
moment map equations split naturally into real and
complex equations associated with $\omega^R$ and
$\omega^C$, the K\"ahler form and
the  holomorphic $(2,0)$ symplectic form, respectively.
The restriction of the \hk metric to the level set is $G$-invariant
and descends to a \hk  metric on the quotient. The quotient
space will be  singular when the group action is not free.

We now specialize the equations to the case of
$p=5,4,3$ branes stuck at a fixed point.

\subsec{Type II}

We begin by considering $X_1=\IC^2$. The relevant
moment maps are:
\eqn\stanmm{
\eqalign{
\mu^C & = [ X, \bar X] + I J \cr
\mu^R & = [X, X^\dagger] + [ \bar X, \bar X^\dagger] +
I I^\dagger - J^\dagger J \cr}
}
for the $U(v)$ gauge group and
\eqn\outmm{
\eqalign{
\tilde \mu^C & = [ Y, \bar Y] + J I \cr
\tilde \mu^R & = [Y, Y^\dagger] + [ \bar Y, \bar Y^\dagger] +
 J J^\dagger -  I^\dagger  I \cr}
}
for the $U(w)$ gauge group.

The moduli space of D-brane ground states (which may
be identified with the classical moduli of the $d=6, \CN=1$
SYM theory on the world-volume) is just the \hk quotient:
\eqn\simpquot{
\mu^C=0,\qquad \mu^R=0\ \mod\ U(v)
}
 together with
$\tilde \mu^C=\tilde \mu^R=0\ \mod\ U(w)$.

Let us now consider the D-branes on an ALE space
$X_n(\vec \zeta)$ realized as a blown-up orbifold.
As we have seen, the gauge symmetry is broken
to
\eqn\brkgg{
\eqalign{
U(w)  & \rightarrow U(\vec w) \equiv U(w_0)  \times \cdots
\times U(w_{n-1}) \cr
U(v)  & \rightarrow U(\vec v) \equiv U(v_0)  \times \cdots
\times U(v_{n-1}) \cr}
}
by the orbifold. The moment maps $\mu_i^R, \mu_i^C$ for
the unbroken gauge symmetry are easily obtained by
substituting the block-diagonal forms of $X,\bar X $
(see \matrep) into \stanmm\ to get
$\vec \mu = Diag\{ \vec \mu_0, \dots, \vec \mu_{n-1} \}$ with:
\eqn\qvrmm{
\eqalign{
\mu_i^C & = X_{i,i+1} \bar X_{i+1,i} - \bar X_{i,i-1} X_{i-1,i} + I_i J_i \cr
\mu_i^R & = X_{i,i+1} X^\dagger_{i,i+1} - X^\dagger_{i-1,i} X_{i-1,i}
+
\bar X_{i,i-1} \bar X^\dagger_{i,i-1} - \bar X^\dagger_{i+1,i} \bar X_{i+1,i}
+ I_i I_i^\dagger - J^\dagger_i J_i \cr}
}
The equations $\mu_i = 0 \mod U(\vec v), \tilde \mu_i=0 \mod U(\vec w)$
give the moduli of $\IZ_n$ equivariant D-brane configurations on
$\IC^2$.

As discussed in the previous section we may -- by turning on
twist fields in the $\sigma$-model of the string theory -- resolve
the target space. Turning on twist fields induces FI terms in
the d=6 SYM theory and the new vacuum equations are
consequently
\eqn\qvrman{
\eqalign{
\mu_i^C & = \zeta_i^C \cr
\mu_i^R & = \zeta_i^R \qquad \mod\ U(\vec v)  \cr}
}
in addition to similar equations for $U(\vec w)$.
The equations \qvrman\  ({\it without}  the $\tilde \mu$ equations)
define what is known as a quiver manifold
\eqn\defquivr{
\CM_{\vec \zeta} (\vec v, \vec w) \equiv
\{ (X,\bar X, I, J): \vec \mu = \vec \zeta \} / U(\vec v)
}
See
\nakalg\ for an extensive discussion of the properties of these manifolds
and for references to the literature. When the action of
$U(\vec v)$ is free on the solutions of \qvrman\ the dimension
of the moduli space is
\eqn\dimform{
\dim_R \CM_\zeta(\vec v, \vec w)=
4 \vec v \cdot \vec w - 2 \vec v \tilde C \vec v
}
The manifolds are generically smooth and topologically
very rich. They do develop important singularities at
nongeneric values of $\vec \zeta$. An extreme case occurs
when $\vec \zeta=0$ and the manifolds are extremely
singular.  Note that, by taking a trace of \qvrmm\ we see
that  compatibility of the D-flatness conditions requires
\eqn\zetacond{
\eqalign{
\sum v_i   \zeta_i^c & = \sum \Tr(I_i J_i) \cr
\sum v_i   \zeta_i^r & = \sum \Tr(I_i I_i^\dagger - J_i^\dagger J_i ) \cr}
}
When this condition is violated there is a potential for the
gravitational moduli $\vec \zeta_i$.

For fixed $Y$ the quiver variety  $\CM_{\vec \zeta}(\vec v, \vec w)$
 describes the
classical moduli of ground states. For $p=5,4$ this is the
same as the quantum moduli. For $p=3$ we can have
an interacting SYM theory, but the hyperkahler metric
is not corrected by quantum effects in   SYM.  The
situation is less clear when including gravitational
interactions. However, $\CM_{\vec \zeta}(\vec v, \vec w)$
is clearly correct
to leading order in the string coupling.

In order to understand better the physical significance of
the quiver varieties let us take $\vec w=0$, i.e., no
outer D-branes at all, and consider, moreover,  a single
D-brane. Such a D-brane must be able to move away from
the orbifold fixed point, and when it does so, it is described
by a symmetrical configuration of $n$ images. Hence,
a single D-brane transverse to an ALE space is described
by $\vec  v = \vec n \equiv ( 1, 1, \dots , 1)$. For such a
choice of $(\vec v, \vec w)$ the action of the group
$G(\vec v) = U(1) \times \cdots \times U(1)$ is not free.
At best $G'(\vec v) = G(\vec v)/ U(1)_{diagonal} $
can act freely. Correspondingly, from
\zetacond, we have the
restriction:
  \eqn\sumzet{
\sum_{i=0}^{n-1}  \vec \zeta_i  = 0
}
on the levels. For $\vec \zeta$ which are otherwise
generic the group $G'(\vec v)$ in fact does act
freely and, by \dimform\  the quotient is a smooth
four-dimensional \hk manifold. In fact, a
theorem of Kronheimer \kronheimer\
asserts that the quiver
variety is the ALE space with periods determined
by $\vec\zeta$ described in section 2.1:
\eqn\krnthm{
\CM_{\vec \zeta}(\vec n,\vec 0) = X_n(\vec \zeta)
}
Thus, Kronheimer's theorem fits in beautifully with the
results of D-brane theory:  The low energy dynamics
of a D-brane transverse to an ALE space may be
described both from a 10 dimensional perspective
and from a 6-dimensional world-volume perspective.
{}From the 10-dimensional viewpoint the dynamics is
clearly given by supersymmetric quantum mechanics
with the ALE space $X_n(\vec \zeta)$ as the target.
{}From the world-brane point of view we have
supersymmetric quantum mechanics with
target space the vacuum manifold
$\CM_{\vec \zeta}(\vec n,\vec 0) $ of  the
$d=6, \CN=1$ SYM. Kronheimer's theorem,
\krnthm\ identifies these as the same target.

This is as much as we have any a priori right to expect,
but, in fact, much more is true: the full dynamics is
described by a sigma model with target
given by $\CM_{\vec \zeta}(\vec n,\vec 0) $. This
suggests a conjecture below.

The interpretation of $\CM_{\vec \zeta}(\vec v, \vec w) $
for other D-brane configurations will be described in
the next section.

\subsec{Type I}

The equations for the type I quivers are
{\it the same} as the equations described in the
previous subsection. The only new point is that the
restrictions on the hypermultiplets described in
sections four and five restricts the moment maps
to take values in the Lie algebras described in
subsection 4.4.

For example, consider case I of subsection
5.1. Using the conditions
\hyphomii\hyphomiii\hrealty\  one can
check that
 \eqn\mtchmm{
\eqalign{
(\epsilon \mu_0^\Lambda)^{tr} & = +\epsilon \mu_0^\Lambda \cr
(\mu_i^\Lambda)^{tr} & = - \mu_{n-i}^\Lambda \qquad 1\leq i \leq n-1,
  i\not= n/2 \cr
(\epsilon \mu_{n/2}^\Lambda)^{tr} & = + \epsilon \mu_{n/2}^\Lambda\cr}
}
Thus, the orientation conditions on the fields already
guarantee that the moment maps take values in the
Lie algebras of the unbroken gauge symmetry.
Similarly, in case II of subsection 5.2 one can check that
\eqn\mtchmmp{
(\mu_i^\Lambda)^{tr}  = - \mu_{n+1-i}^\Lambda \qquad 1\leq i \leq n
}
in accord with the relevant
gauge groups.
Note that a rather peculiar feature
of  \mtchmm\   is that  the equations \qvrman\
only make sense for
\eqn\symmcond{
\vec \zeta_i = - \vec \zeta_{n-i}  \quad .
}

Further discussion of Type I quivers will be found in
\dmns.

\newsec{Type II D-branes and $U(N)$ gauge instantons on ALE }

\def\sqm{supersymmetric quantum mechanics\ }

Let us consider two {\it a priori} distinct physical
situations. For definiteness we focus on the
\IIb\ theory.
First we consider $w$ type \IIb\ 7-branes whose
world-volume is given by the supersymmetric
cycle $\CB_7 = \IR^3 \times X_n(\vec \zeta)$.
Semiclassical supersymmetric groundstates of this
7-brane theory will be described by $U(w)$ instantons.
In particular, let us focus on the low energy
dynamics of the  states given by instantons on
$X_n(\vec \zeta)$.  The low energy dynamics of
these states is described by \sqm
with the target being the moduli of
$U(w)$ instantons $\CM_{inst} $.

Next,  we consider the supersymmetric boundstates of
3-branes  $\CB_3 = \IR^3 \times \{ P_0\} $
where $P_0\in X_n(\vec \zeta)$ is a point with the
7-branes $\CB_7$. The
moduli  of these latter boundstates can be given an explicit
description using the orbifold construction of this paper.
Namely, we consider $w$ 7-branes on
$\IR^3 \times \IC^2/\Gamma$ and $v$ 3-branes on
$\IR^3 \times \{ 0 \} $. Turning on FI terms
$\vec \zeta$, restricted by the condition
\sumzet, resolves the $\sigma$-model on the
orbifold to the sigma model on $X_n(\vec \zeta)$
as discussed in the previous section. At the same
time, turning on $\vec \zeta$ resolves the moduli
of 3-brane collective coordinates: The
low energy dynamics of the 3-brane degrees of
freedom, for fixed 7-brane degrees of freedom,
is governed by \sqm with target
$\qmvw$.
%
%

According to the results of \witten\douglas, these
two \sqm\  systems should be the same. Thus
we expect that the moduli of instantons
on $X_n(\vec \zeta)$ should be identified
with $\qmvw.$ To be a little  more precise,
if the instanton breaks
\eqn\instbrsk{
U(w) \rightarrow U(w_0) \times \cdots \times U(w_{n-1})
}
at infinity in $X_n(\vec \zeta)$ then we should
identify such instantons with the $(3,7)$ branes
where the 7-branes also induce the same
breaking \instbrsk.
In fact, the identification of
$\CM_{inst}$ with $\qmvw$ is   known to be
correct and is called the Kronheimer-Nakajima
theorem.  In the next two sections we
describe the identification more precisely.
\foot{
Moreover, given the case
$W=0$ it is natural to conjecture
that in fact the exact dynamics is described as
a string theory with the moduli space as target. }

\subsec{Topological classification of
instantons on $X_n(\vec\zeta)$. }

We must first describe the data needed to
specify the components of the moduli spaces.
The space of $U(w)$ connections on a complex
$w$-dimensional vector bundle $E\rightarrow X_n$ of
finite action is classified, topologically,  by
$c_1(E)\in H^2(X_n;\IZ)$, $ch_2(E)$,  and
a homomorphism $\rho: \IZ_n \rightarrow U(w)$.
The last piece of data corresponds to the data
specifying a flat connection at infinity.
If $\rho$ is conjugate to $g \rightarrow
Diag\{ \xi^j 1_{w_j} \}$ then we may
equivalently say that the unbroken
gauge group at infinity is $\prod U(w_j)$.
We will denote the space of  $U(w)$ instantons
on $X_n$ by
$\CM(X_n(\vec \zeta), U(w) ;c_1,ch_2,\rho)$.

We will need to be more explicit about the
nature of $c_1$.
We first introduce a set of line bundles
$\CR_i$.  The construction of $X_n$ as
a \hk\ quotient identifies it with a
principal
$\prod_{i=1}^{n-1}U(1)$ bundle.
Choosing the associated bundle
with charge $1$ for the $i^{th}$ $U(1)$
defines $\CR_i$. These line
bundles carry a natural connection
such that $c_1(\CR_i)$
are harmonic 2-forms forming  a
basis for $H^2$ dual to the basis $\Sigma_i$
of $H_2(X_n;\IZ)$.
Thus we may expand:
\eqn\chrni{
c_1(E)  = \sum_{i=1}^{n-1} u^i c_1(\CR_i)
}
for some vector $\vec u\in \IZ^{n-1}$.

It can be shown that
\eqn\crtii{
c_1(\CR_i)\cdot c_1(\CR_j) \equiv
\int_X  c_1(\CR_i)\wedge c_1(\CR_j)= - (C^{-1})_{ij}
}
where $C$ is the Cartan matrix.
In particular,  from \crtii\ we see that:
\eqn\chrniii{
ch_2(\CR_i) =  \half {i (n-i)\over  n}
}
This can be fractional because $X_n$ is noncompact.

\subsec{The Kronheimer-Nakajima Theorem}

We now come to the remarkable Kronheimer-Nakajima
theorem \kn\  which gives the
isomorphism of \hk manifolds:
\eqn\knthrm{
\CM_{\vec\zeta}(\vec v, \vec w)
\cong
\CM(X_n(-\vec \zeta), U(w) ;c_1,ch_2,\rho)
}
where $(\vec v, \vec w)$ are related to
topological  quantities $c_1(E), ch_2(E), \rho$
as follows.
At in a neighborhood of infinity
we may think of the bundle
$E_\infty \cong \oplus w_i \CR_i$
where $\CR_i$ are flat line bundles
on $S^3$ associated to the $i^{th}$ representation
of $\IZ_n$.
The first Chern class is obtained from \chrni\ with
$\vec u = \vec  w - \tilde C \vec v$, where
$\tilde C$ is the extended Cartan matrix.
\foot{Curiously, the  $\beta$ function of $SU(v_k)$ is
$\beta(SU(v_k)) =  {u_k \over  16 \pi^2} $. That is,
the beta function is the first Chern class!  Moreover,
it can be shown that $\CM_{\vec 0}$ has regular
points only when $u_k \geq 0$ \nakalg. This coincidence
should have a simple physical explanation.
}
Finally $ch_2(E)$ is given by:
\eqn\chernii{
ch_2(E) = \sum_{i=0}^{n-1} u_i ch_2(\CR_i) + {1\over  n} \dim V
}
These equations have a beautiful and simple
interpretation: The first term comes from magnetic monopoles
centered in the different exceptional divisors $\Sigma_i$.
Far away from these divisors $X_n(\vec \zeta)$ appears
like $\IR^4/\IZ_n$. Instantons on this space look like
ordinary $\IZ_n$-invariant instantons on $\IR^4$.
Alternatively, D-branes carry instanton charge $1/n$.
Some further manipulation leads to a useful alternative formula:
\eqn\othcher{
ch_2(E) = v_0 + \sum_{i=1}^{n-1} {i(n-i) \over  2n} w_i
}
Note that if we fix $w_i$ and vary $v$ then only
$v_0$ contributes to the instanton number.

Remarks:

\item{1.} In fact, from ADHM data we can reconstruct $E$ and
the gauge field quite explicitly \kn. Some explicit
examples of the construction can be found in  \bianchi.

\item{2.} The extra minus sign in \knthrm\  is surprising.
However recall that
$X_n(\vec \zeta) \cong X_n(-\vec \zeta)$.

\item{3.}  Thus far we have been assuming the
condition \sumzet.  For general $(\vec v, \vec w)$
this is not a generic choice of $\vec \zeta$.
Thus, the instanton moduli
space will have some singularities.
These are associated to zero-scalesize limits.

\item{4.} The generalizations of these statements to
$SO(w)$ instantons via Type I theories will be
discussed in \dmns.

\subsec{Torsion free sheaves}

In the previous sections we have interpreted D-brane
moduli spaces $\qmvw$ under the condition
\sumzet. In general, when $\vec w \not=0$
this condition is unnecessary.
\foot{For type \I\ it follows from \symmcond.}
For example, on $X_1=\IR^4$ we can
  have $\vec \zeta_0 \not=0$ by giving
vacuum expectation values to the
self dual parts of $B_{mn}$.

Introducing levels with $\sum \vec \zeta\not=0$
changes two aspects of our understanding of
the collective coordinate space.

First, $\qmvw$ has  singularities  when
\sumzet\  holds but  in general is smooth.
Indeed  one can define a map
\eqn\ressings{
\qmvw \rightarrow \CM_{\vec 0} (\vec v, \vec w)
}
which is a resolution of singularities
\nakalg\nakresol\nakheis. Physical
mechanisms often smooth out the
singularities of moduli spaces and this
is yet another example.

Second, when $\sum \vec \zeta\not=0$,
$\qmvw$ can still be interpreted as a
moduli space of geometric objects,
which generalize Yang-Mills instantons.
By the Donaldson-Uhlenbeck-Yau
theorem (which continues to hold
on the noncompact spaces $X_n(\vec \zeta)$)
the moduli of  ASD instantons can be
identified with the moduli of holomorphic
vector bundles. When
$\sum \vec \zeta\not=0$ we must generalize
the notion of holomorphic
vector bundle to that of torsion free
sheaves.  Without going into technical
details this means -- roughly --
 that the rank of the fiber can change at
isolated points. We must introduce
extra  pointlike degrees of freedom.

For example, on $X_1$,
$\CM_0(v,w)$ is the moduli of
$U(w)$ instantons of instanton number
$v$. This space is singular because
the scale size of an instanton can
shrink to zero. By contrast for
$\zeta\not=0$,
$\CM_\zeta(v,w)$
is the moduli of rank $w$ torsion
free sheaves $\CS$ with
$ch_2(\CS)=v$ on
$\IP^2$. As an indication of how
different these spaces can be note
that for $w=1$, there are no nontrivial
line bundles on $X_1$ so
$\CM_0(v,1)$ is a point. However,
$\CM_{\zeta}(v,1)$ is isomorphic
to the Hilbert scheme of points $[\IC^2]^{[v]}$.
For details and more
precise statements see \nakresol\nakheis.
This example generalizes to the other
ALE spaces.
Note that \ressings\ may be interpreted as
saying that $\qmvw$ is a smooth \hk
compactification of the moduli of instantons.

The generalization from vector bundes
to sheaves is significant for several
reasons. First, it indicates an important
conceptual change since it introduces
a wider and more flexible class of geometric
objects in string compactification.
Second, it has been observed
\nakheis\grojn\lmns\ that the appearance
of infinite dimensional algebras in
the context of gauge theories,
discovered by Nakajima
\nakalg,  necessitates the generalization
of vector bundles to sheaves.
Moreover,  these algebras are related -
in ways not yet clearly understood -
to duality symmetries in string theory
and supersymmetric field theory.
Third,  several recent calculations
\vw\sen\vafadb
\ref\bsv{ M. Bershadsky, V. Sadov, and
C. Vafa, ``D-Branes and Topological Field Theories,''
hep-th/9511222;
``D-Strings on D-Manifolds,''
hep-th/9510225}
\stromvaf\
involving
the counting of D-brane bound states
have used the smooth hyperk\"ahler
resolution of instanton moduli space.
The use of torsion free sheaves as a
resolution of singularities of instanton
moduli spaces deserves to be understood
much better. The extra degrees of
freedom  somehow resolve the boundary
of instanton moduli space, and these
degrees of freedom are crucial to the
counting of BPS states used in verifying
predictions of duality in
\vw\sen\vafadb\stromvaf.

Finally, we remark that there is one important
qualitative difference between the modulus
$\sum \vec \zeta$ and the remaining moduli.
Since the self-dual forms on $X_n$ are
not normalizable, the modulus
$\sum \vec \zeta$ does not fluctuate, in
contrast to the remaining degrees of
freedom in $\vec \zeta$.

 \newsec{Conclusions}

Strings can be sensibly compactified on certain singular spaces, producing
completely non-singular effective field theories.
Orbifolds were the first example of this --
not only are they non-singular CFT's, but turning on twisted sector moduli
resolves the fixed points and produces a smooth manifold, verifying that
these are limits of smooth manifolds with singular metric but non-singular
physics.

In this work we showed that a straightforward treatment of D-branes on an
orbifold reproduces the region of moduli space
around this singular point in a simple
and mathematically natural way.
Furthermore, the D-branes provide descriptions of instanton moduli spaces.

We worked with a non-compact target space,
an orbifold containing a single $\BZ_n$ singularity,
which when resolved becomes an ALE space.
Moduli spaces of metrics and instantons on ALE spaces were constructed as
\hk\ quotients by Kronheimer and Nakajima \refs{\kronheimer,\kn} and the
D-brane theory (for $U(n)$ gauge groups) reproduces their construction
exactly.
It is straightforward to get an explicit metric on the ALE spaces and moduli
spaces
from this construction, and indeed this has already been done in
\refs{\kronheimer,\kn}.

The qualitative structure of moduli spaces for compact target spaces obtained
by resolving orbifolds,
and in particular the enhanced gauge symmetry of the zero instanton
size limit, will be determined by the behavior at the orbifold singularities.
To find consistent models one must implement the tadpole conditions of
\gimon.
It would be quite amusing if the solutions included
models with fixed points of high order.
The rank of the enhanced gauge symmetry for a model with $k$ $5$-branes
at a $\BZ_n$ fixed point is roughly $r\sim kn$ (for type \II;
$r \sim kn/2$ for type \I).
Taking at face value the possibility of shrinking all instantons in a 
compactification with $p_1(V)=p_1(TM)$, one might
have $k=p_1$, while in six dimensions $\BZ_{12}$ orbifolds exist
(and even higher order singularities on less symmetric manifolds),
so it is conceivable that extremely large gauge groups can be obtained.

In string theory, these results are corrected at all orders in $1/\alpha'$
and the string coupling $\lambda$.
By extending the world-sheet computation of the FI couplings as described
in the appendix, it may be possible to obtain the exact (in $\alpha'$)
relation between twist field moduli $\phi$
and periods $\zeta$, and the exact metric on moduli space.
Although we expect corrections in the string
coupling as well, using duality to combine these results with the known
results for the heterotic string may allow controlling them.

In a sense, this model describes a change of topology on
the {\it microscopic} level.  Many examples of topology change are known,
some involving D-branes, but so far the arguments are based on
properties of the low-energy effective theory.  In our example we have
a complete microscopic theory realizing a very simple topology change,
the resolution of a singularity.  Perhaps other changes of topology can
be similarly realized, helping to provide insight into the concepts which
must replace metric and topology in a complete theory.

\subsec{Reciprocity and T-duality}

Finally, let us point out one very interesting
direction for future research.~\ref\newstuff{
M. Douglas, G. Moore, N. Nekrasov and S. Shatashvili, work in progress.}
The quiver diagrams,
figs. 5-11 , exhibit an intriguing symmetry:
One can switch the inner quiver for the outer quiver.
This corresponds to exchanging, say, 5-branes and
9-branes and hence corresponds to $T$-duality.
It follows from the instanton interpretation that
there should be a duality between $U(w)$ instantons
of charge $v$ and $U(v)$ instantons of charge $w$, etc.
For example, we expect that applying $T$-duality
to configurations of  $(w,v)$ $(9,5)$-branes
on $T^4 \times \IR^{5,1}$ with the 5-branes
transverse to a torus $T^4$ gives a
stringy realization (and generalization)
 of the Fourier-Nahm-Mukai
transform
\ref\nahm{W. Nahm, ``Monopoles in quantum field
theory,'' in {\it Proceedings of the Monopole
Meeting, Trieste}, Craigie et. al. eds, World Scientific;
``Self-dual monopoles and calorons,'' in
Lecture notes in physics, vol. 201, G. Denardo et. al.,
eds. Springer 1984}
\ref\corrigan{E. Corrigan and P. Goddard,
``Construction of Instanton and Monopole solutions and
reciprocity,'' Ann. Phys. {\bf 154} (1984)253}
\ref\shenk{H. Shenk, ``On a generalized Fourier
Transform of Instantons over flat tori,'' Commun.
Math. Phys. {\bf 116}(1988)177}
\ref\braam{P.J. Braam and P. Van Baal, ``Nahm's
Transformation for Instantons,'' Commun. Math. Phys.
{\bf 122}(1989) 267}
\ref\donkrn{S.K. Donaldson and P.B. Kronheimer,
{\it The geometry of four-manifolds} Oxford, 1990}.
It is straightforward to check that the mapping of
the Chern classes \braam\  is precisely the
mapping of RR charges predicted by  T-duality.

 A crucial role in the discussion of  instanton
reciprocity   (and of the proof of completeness
of the ADHM construction)  is played by the Dirac
operator in the field of an instanton.
Significantly, the DN fields are valued in the spinor
bundle on $X_n(\vec \zeta)$. Indeed, letting
$m=1, \dots v$ label a basis of zeromodes of the
Dirac operator and $M=1,\dots, w$ denote
indices with respect to a basis of $E_\infty$,
 the asymptotic
behavior of a solution of the Dirac equation in the
field of an instanton is
\eqn\dirac{
\psi^{M}_{~~mB}(x) \sim
\tilde h^{AM}_{~~m} {x_{AB} \over  x^4}
}
in Euclidean coordiantes $x$. We hope the  connection
\dirac\  to fields of a string theory
will lead to a deeper understanding of the
completeness of the ADHM construction and of
instanton reciprocity.

\bigskip
\bigskip

\centerline{\bf Acknowledgements}

We would like to thank the ITP at Santa Barbara
for providing a stimulating atmosphere in which
this research was initiated.  We would
like to thank  M. Berkooz, J. Harvey,
R. Leigh, A. Losev, H. Nakajima,
 N. Nekrasov, J. Polchinski, A. Strominger,
J. Schwarz,  A. Sen, and
S. Shatashvili for useful discussions and correspondence.
GM thanks the Rutgers high energy theory
group for hospitality while this research was
concluded.
The research of GM
is supported by DOE grant DE-FG02-92ER40704,
and by a Presidential Young Investigator Award,
while that of MRD was supported in
part by DOE grant DE-FG05-90ER40559 and NSF
PHY-9157016.

\appendix{A}{World-sheet computation of twist field couplings}

Here we verify the results \csi\ and \ggstsum\ of section 6,
by world-sheet computation.
These come from the two-point function
\eqn\fcoup{\vev{V(C^{(4)}_k)  V(F)}}
and the integrated three-point functions
\eqn\xcoup{\eqalign{
\vev{V(\phi_k^{CD}) V(X^{A'A}) \int V(X^{B'B})  }\cr
\vev{V(\phi_k^{CD}) V(H^{A}) \int V(H^{B}) }
}}
on a disk with boundary on the D-brane, $F$, $\Psi$ or $H$ integrated over
the boundary, and $\phi$ in the interior.  In fact, we will only
check the cross term in \poteng. (The remainder is determined by
supersymmetry.)

The amplitude factorizes into three parts: the classical action of the
relevant embedding of the disk into the target space, the quantum CFT
amplitude, and the contribution of the Chan-Paton factors.
If the D-brane is coincident with the fixed point, the embedding
is trivial.  More generally, the twist operator $V(H)$ will
constrain the embedding to the fixed point at the insertion,
while the Dirichlet boundary conditions are fixed at the position
of the D-brane, so that the embedding will be non-trivial.
The classical action will be proportional to the distance
squared, leading to an overall factor $\CA \propto \exp -|X|^2/\alpha'$.
However, we will not consider this case in further detail here, instead
working in the limit $|X|^2<<\alpha'$.

We now explain the definition of Chan-Paton factors in twisted sectors.
First, in an untwisted sector, we sum over all orderings of the
operators on the boundary, and include the trace over their Chan-Paton
matrices $\lambda$ with the same ordering:
\eqn\defordcp{\eqalign{\CA =
\sum_{\sigma\in S_N} \int^{2\pi}_0 d\theta_{\sigma(1)}\ %
&	\int^{\theta_{\sigma(1)}}_0 d\theta_{\sigma(2)}\ \ldots
	\int^{\theta_{\sigma(N-1)}}_0 d\theta_{\sigma(N)}\cr
&	\tr \lambda_{\sigma(1)}\ldots\lambda_{\sigma(N)}
	\vev{ V_1(\theta_1)\ldots V_N(\theta_N)}.
}}
In a sector twisted by $g$, the contribution of the Chan-Paton factors is
modified to be
\eqn\twiscon{
\tr \gamma_g\lambda_{\sigma(1)}	\ldots\lambda_{\sigma(N)} }

In defining the action of the twist on fields such as $X^{A'A}$ with
internal Lorentz indices, we have been assuming that the twist can
act both on Chan-Paton and Lorentz indices (as in \orbrot), and that
the projection retains the singlet under the combined action.
While this definition is certainly intuitive, we should check that it is
consistent.  What makes this less than obvious
is that we will be using operators with multi-valued world-sheet correlation
functions.  For example, in \xcoup,
the operator $X$ transforms as an internal vector,
so in circling a $g^n$ twist field, its correlators
will have monodromy $\xi^n$.
We need to understand in what sense the twist on the Chan-Paton
factors can compensate for this.

The point will be to associate the cuts in a correlation function
with the twist $\gamma_g$.
We thus choose a location $\theta_g$ for the twist on the boundary,
and define correlators to be single-valued except on a cut from the twist field
to $\theta_g$.
For example, to reproduce \twiscon\ from the definition \defordcp,
we should let $\theta_g=2\pi$.

The sum in \defordcp\ is now over all orderings of the vertex operators
and $\gamma_g$.  Since the Chan-Paton matrices $\lambda(X)$ for the operators
$X$ do not commute with $\gamma_g$, all orderings can contribute to
different correlation functions.  In our calculations,
$\lambda(X_{k,k+1})\gamma(g^n)=\xi^n\gamma(g^n) \lambda(X_{k,k+1})$,
and this difference will be reflected in a phase.  After the discrete
Fourier transform of subsection 6.4, it is clear that these two amplitudes
couple to the two different $U(1)$ factors $k$ and $k+1$ under which
$X_{k,k+1}$ is charged.

We should check that the particular value of $\theta_g$ does not appear in
the final result.  If we change it to $\theta'_g$, we will modify
the amplitude by taking the integral over $[\theta_g,\theta'_g]$
away from one correlator (say $\tr \gamma_g\lambda_1\lambda_2$) and
adding it, with the appropriate phase produced by crossing the cut,
to another correlator (say $\tr \lambda_1\gamma_g\lambda_2$).
But the phase produced by the cut will exactly compensate the phase
produced by reordering the Chan-Paton factors, and this will contribute
to the same correlator.

For our purposes, this discussion suffices, assuming that these theories
are sensible.
This remains to be proven by checking factorization
and other consistency conditions.
As we will see below, the use of multi-valued operators adds new elements
to the discussion, and such a proof is an important open problem.

We proceed to the world-sheet calculation. The $SL(2,\BR)$ conformal
symmetry of the disk can be used to fix the positions of $V(\phi)$, $V(C)$ and
one boundary operator.
We conformally transform the disk to the upper half plane and
work on its double $\BC$, mapping the boundary to the real axis.
The Dirichlet boundary conditions are encoded in simple
transformation properties for the fermions.
We split the operator $V(\phi)$ into its original chiral part at $z=i$ and
the mirror of its anti-chiral part at $z=-i$.

\subsec{Chern-Simons couplings}

We need to choose a picture for the vertex operators compatible with the
total superconformal ghost number $-2$ of the disk.
For \fcoup, since the $(-1/2,-1/2)$ picture for $V(C)$ is by far the
simplest, we use the $-1$ picture $V(A_{\mu})=\psi^\mu~e^{ikX}~e^{-\phi}$,
deriving the equivalent coupling
\eqn\cscouplg{
-\int A\wedge H^{(d-1)}
}
Recall that for  these we have the worldsheet correlator
(for a \IIb theory):
\eqn\wscoup{
\langle S_\alpha(z) \psi^\mu(x)   (\Gamma^{6789})_\beta^{~\rho} S_\rho(\bar z)
\rangle_{\IP^1}
= {   (\Gamma^{\mu 6789}C)_{\alpha\beta}\over
(z-x)^{1/2} (x-\bar z)^{1/2} (z-\bar z)^{3/4} }
}
leading to the untwisted coupling
$\int \Tr A_\mu H_{\alpha \beta} (\Gamma^{\mu 6789} C)^{\alpha \beta}$
where the trace is on Chan-Paton indices.
For the twisted case the calculation is essentially
the same as for untwisted couplings of this type.
The twist eliminates the fermion zero modes in the internal space,
(whether or not the brane is transverse),
so the R-R boundary state is a bispinor in $d=6$.
$V(A_{\mu})$ is trivial in the internal space, so the calculation simply
reduces to the untwisted calculation in $d=6$.
This depends trivially on the dimension $p$ of the D-brane (e.g. see \li).

\subsec{FI - couplings}

Some but not all of the amplitudes \xcoup\ are determined by supersymmetry.
We do $\vev{\phi X X}$, the other one is similar.
On symmetry grounds we must find
\eqn\coupgroup{
\vev{V(\phi_k^{CD}) V(X^{A'A}) \int V(X^{B'B}) }\sim
\eps^{A'B'} (a \eps^{AB} \eps^{CD} + b \eps^{A(C}\eps^{D)B} )
.}
The coupling $b$ is \ggstsum, expected by supersymmetry, but the
coupling $a$ is unrelated by world-brane supersymmetry and directly couples
the invariant $X\bar X$ to the singlet in \twisty\ present
in the type \II\ strings (corresponding to the integral of $B$ about a
two-cycle).  Supersymmetry does not appear to be
compatible with $a\not=0$.
%
%

The NS-NS twist field is simplest in the $(-1,-1)$ picture,
so we take the $V(X)$'s in the $0$ picture.
Now we need explicit twist field correlators.
Fermi correlators can be computed via bosonization; e.g. write
$\psi^{A'}=e^{iH_{A'}}$ and $\bar\psi^{A'}=e^{-i H_{A'}}$, (for left movers;
resp. $\tilde H$ for right movers).
Then one twisted massless state is
$e^{iH_2+i\tilde H_1}~ e^{ik(H_1-H_2+\tilde H_2-\tilde H_1)/n}\ket{0}$.
Bose correlators can be done as in \orbcft.
The results we need
(on the sphere and up to an overall function of $z_1-z_2$) are
\eqn\twistcor{\eqalign{
\eps_{C'D'}\langle &\sigma_k\psi^{C'C}(z_1)\psi^{A'A}(x_1)\psi^{B'B}(x_2)
	\psi^{D'D}\sigma^\dagger_k(z_2) \rangle \cr
	&= \eps^{A'B'}
		\left({z_1-x_1\over z_2-x_1}\right)^{(-)^{A'}k/n}
		\left({z_1-x_2\over z_2-x_2}\right)^{(-)^{B'}k/n} \times\cr
&\qquad		\left[{\eps^{AB}\eps^{CD}\over (z_1-z_2)(x_1-x_2)}+
			{\eps^{AC}\eps^{BD}\over (z_1-x_1)(z_2-x_2)} +
			{\eps^{AD}\eps^{BC}\over (z_1-x_2)(z_2-x_1)}
		\right]
}}
\eqn\twistcorb{\eqalign{
\langle \sigma_k&\p X^{A'A}(x_1)\p X^{B'B}(x_2) \sigma^\dagger_k(z_2) \rangle
\cr
	&= \eps^{A'B'}\eps^{AB}
		\left({z_1-x_1\over z_2-x_1}\right)^{(-)^{A'}k/n}
		\left({z_1-x_2\over z_2-x_2}\right)^{(-)^{B'}k/n} \times\cr
&\qquad		{1\over (x_1-x_2)^2}
		\left[\left(1-{k\over N}\right) + {k\over N}
			\left({z_1-x_1\over z_2-x_1}\right)^{-(-)^{A'}}
			\left({z_1-x_2\over z_2-x_2}\right)^{-(-)^{B'}}
		\right].
}}

A general correlation function is
\eqn\reprecor{\eqalign{
&\vev{V(\phi_k^{CD})  V(X^{2B}) \int V(X^{1A})}\cr
&= 2\int dx_1\ %
 \langle
 c\gamma \sigma_{k} \psi^{1C}e^{\half ip_3 X }(z)
 ~(i \psi^{1A} p_1\cdot\psi+\p X^{1A})e^{ip_1 X}(x_1)\cr
&\qquad\qquad~~ c (i \psi^{2B} p_2\cdot\psi +\p X^{2B})e^{ip_2 X}(x_2)~
 c\gamma \psi^{2D}\sigma^\dagger_{k}  e^{\half ip_3 X}(\bar z) \rangle.
}}
As in \dis, we will need to take the momenta slightly off-shell:
we can take $\delta\equiv p_1\cdot p_2$, $p_3\cdot p_1=p_3\cdot p_2=-\delta$.

Evaluating the correlation function at $z=i, x_2=0$ and $x_1=x$ gives
\eqn\evrecor{\eqalign{
\int_{0}^\infty &{dx}|i-x|^{-\delta}x^{\delta}\ %
		\left({x-i\over x+i}\right)^{(-)^{A'}k/n}
		(-1)^{(-)^{B'}k/n} \times\cr
&\qquad		\bigg[{\delta\over x}\left(
			{\eps^{AB}\eps^{CD}\over 2ix}+
			{\eps^{AC}\eps^{BD}\over i(x-i)} -
			{\eps^{AD}\eps^{BC}\over i(x+i)}
			\right) - \cr
&\qquad\qquad\qquad	{\eps^{AB}\eps^{CD}\over 2ix^2}
			\left(
			  \left(1-{k\over N}\right) - {k\over N}
			  \left({x-i\over x+i}\right)^{-(-)^{A'}}
			\right)
		\bigg]	.
}}
The integral could produce poles as $x\rightarrow 0$ for $\delta=1,0,\ldots$.

Let us first consider the triplet coupling to $\vec\sigma_C^D$.
This reduces the expression in square brackets to
$[2\delta/x(1+x^2)]$. Doing the integral one finds integral has a pole at
$\delta=0$,
cancelling the factor of $\delta$. The $\delta \rightarrow 0$ limit is
simply equal to $2$,  and in particular is independent of $k/n$.

The singlet coupling for $k=0$ is also easy:
the terms in the brackets combine to $[(\delta-1)/x^2]$.
The integral is facilitated by changing variables $x^2=y$,
producing $(\delta-1)\Gamma((\delta-1)/2)\Gamma(1/2)/\Gamma(\delta/2)$.
The prefactor $(\delta-1)$ cancels the pole at $\delta=1$, and the
result is zero for $\delta=0$.

The computation for $k\ne 0$ is substantially more difficult,
and we only give the highlights.  The double pole cancels as for $k=0$;
and there is no single pole (by the definition of the bosonic correlator).
Getting the finite part requires doing the integral.
We took the cuts in the $\Im x<0$ half-plane, and rotated the contour
to the positive imaginary axis.  The branch point at $x=i$ requires
dividing the contour into two parts and special care with the phases.
A change of variables $x=it/(1-t)$ turns these into $t\in[0,\half]$ and
$t\in[\half,1]$. Finally, the integrand can be rearranged into the form
$\int (1-2t)^{-\delta/2}(d/dt)[\ldots]$, which is tractable.

The result is that the two contours combine to zero,
with an appropriate choice of relative phase.
One way to obtain this choice
is to define the correlation function using `radial operator ordering.'
This will turn the cut in the $\vev{\psi\sigma\sigma}$ correlation functions
into a cut at $|x|=1$ with $\psi(1+\eps)=e^{\pi i k/n}\psi(1-\eps)$.

Without a complete analysis of world-sheet consistency it is not
possible to prove that this is the only sensible choice.  However,
as we mentioned earlier, this coupling would not have been supersymmetric,
which is the best argument for its vanishing.  We give this calculation more
as an illustration of a subtlety in the world-sheet
definition of open strings on orbifolds which should be better understood.

\bigskip

\listrefs
\end